\documentclass[aps,pra,twocolumn,amsmath,amssymb,nofootinbib,superscriptaddress]{revtex4}
\usepackage{times}
\usepackage[pdftex]{graphicx}
\usepackage{dcolumn}
\usepackage{bm}
\usepackage{amsmath}
\usepackage{indentfirst}
\usepackage{float}
\usepackage[colorlinks]{hyperref}
\usepackage[dvipsnames]{xcolor}

\usepackage{makecell}
\usepackage[normalem]{ulem}

\newcommand{\Dop}{\mathcal{D}}
\newcommand{\Hop}{\hat{H}}

\newcommand{\Lop}{\mathcal{L}}
\newcommand{\rhoop}{\hat{\rho}}

\newcommand{\aop}{\hat{a}}
\newcommand{\adop}{\hat{a}^{\dagger}}
\newcommand{\paras}{\vec{\alpha}}
\newcommand{\xop}{\hat{x}}
\newcommand{\nop}{\hat{n}}
\newcommand{\nave}{\bar{n}}
\newcommand{\nss}{\bar{n}^{{\rm st}}}
\newcommand{\loss}{{\rm loss}}
\newcommand{\hc}{{\rm H.c.}}
\newcommand{\trace}{{\rm tr}}
\newcommand{\im}{{\rm i}}

\newcommand{\gcc}[1]{{\color{black}{#1}}}

\begin{document}

\title{Fast Laser Cooling Using Optimal Quantum Control}

\author{Xie-Qian Li}
\thanks{These authors contribute equally to this work.}
\affiliation{Department of Physics, College of Liberal Arts and Sciences, National University of Defense Technology, Changsha 410073, China
Interdisciplinary Center for Quantum Information, National University of Defense Technology, Changsha 410073, China}

\author{Shuo Zhang}
\thanks{These authors contribute equally to this work.}
\affiliation{Henan Key Laboratory of Quantum Information and Cryptography, Zhengzhou,
Henan 450000, China}

\author{Jie Zhang}
\affiliation{Department of Physics, College of Liberal Arts and Sciences, National University of Defense Technology, Changsha 410073, China
Interdisciplinary Center for Quantum Information, National University of Defense Technology, Changsha 410073, China}

\author{Wei Wu}
\affiliation{Department of Physics, College of Liberal Arts and Sciences, National University of Defense Technology, Changsha 410073, China
Interdisciplinary Center for Quantum Information, National University of Defense Technology, Changsha 410073, China}

% \author{Wan-Su Bao}
% \affiliation{Henan Key Laboratory of Quantum Information and Cryptography, Zhengzhou,
% Henan 450000, China}

\author{Chu Guo}
\email{guochu604b@gmail.com}
\affiliation{Henan Key Laboratory of Quantum Information and Cryptography, Zhengzhou,
Henan 450000, China}

% \affiliation{Key Laboratory of Low-Dimensional Quantum Structures and Quantum Control of Ministry of Education, Department of Physics and Synergetic Innovation Center for Quantum Effects and Applications, Hunan Normal University, Changsha 410081, China}

\author{Ping-Xing Chen}
\email{pxchen@nudt.edu.cn}
\affiliation{Department of Physics, College of Liberal Arts and Sciences, National University of Defense Technology, Changsha 410073, China
Interdisciplinary Center for Quantum Information, National University of Defense Technology, Changsha 410073, China}

\begin{abstract}
Cooling down a trapped ion into its motional ground state is a central step for trapped ions based quantum information processing. State of the art cooling schemes often work under a set of optimal cooling conditions derived analytically using a perturbative approach, in which the sideband coupling is assumed to be the weakest of all the relevant transitions. As a result the cooling rate is severely limited. Here we propose to use quantum control technique powered with automatic differentiation to speed up the classical cooling schemes. We demonstrate the efficacy of our approach by applying it to find the optimal cooling conditions for classical sideband cooling and electromagnetically induced transparency cooling schemes, which are in general beyond the weak sideband coupling regime. Based on those numerically found optimal cooling conditions, we show that faster cooling can be achieved while at the same time a low average phonon occupation can be retained.
\end{abstract}

\date{\today}
\pacs{}
\maketitle

\address{}

\vspace{8mm}

\section{Introduction}\label{sec:introduction}
Laser cooling of a trapped ion is a central step for coherent manipulation of the underlying quantum state~\cite{WinelandMeekhof1998}, which is vital in various applications such as quantum simulation~\cite{PorrasCirac2004b,PorrasCirac2004a,LeibfriedWineland2003,BermudezPlenio2013,RuizCampo2014,RammHartmut2014,GuoPoletti2015,GuoPoletti2016,GuoPoletti2017,GuoPoletti2017b,GuoPoletti2018,XuPoletti2019,PanDavidson2020,WuChen2019} and quantum computing~\cite{CiracZoller1995,LanyonRoos2011,KielpinskiWineland2002}. Till now the most widely used cooling schemes in experiments include sideband cooling and dark-state cooling
% such as the electromagnetically induced transparency (EIT) cooling~\cite{MorigiKeitel2000, RoosBlatt2000},
due to their efficacy and simplicity for experimental realization.

Sideband cooling scheme can be understood theoretically by expanding the interacting Hamiltonian of the laser and ion to the first order of the Lamb-Dicke parameter, denoted as $\eta$, which would produce three terms: a zeroth order carrier transition between the inner states of the ion, a first order blue sideband in which the ion gets excited and absorbs one vibrational quantum number (phonon), and a first order red sideband in which the ion gets excited and emits one phonon~\cite{DiedrichWineland1989,CiracZoller1992,MonroeWineland1995,RoosBlatt1999}. The red sideband together with the spontaneous emission of the excited state essentially induce the cooling, while the other two transitions induce heating. Dark-state cooling schemes in general involve more internal energy levels as well as more lasers, but could often be understood similarly to sideband cooling by transforming the internal degrees of freedom into appropriate dressed state pictures~\cite{MorigiKeitel2000,RoosBlatt2000,EversKeitel2004,RetzkerPlenio2007,CerrilloPlenio2010,AlbrechtPlenio2011,ZhangChen2012,YiYang2013,ZhangChen2014,LuGuo2015,YiGu2017,CerrilloPlenio2018,ZhangGuo2021b}. Compared to sideband cooling, dark-states cooling schemes often have the advantage that the carrier transition, or the blue sideband, or both of them get eliminated in the first order of $\eta$. However, both sideband cooling and dark-state cooling schemes derive the optimal cooling conditions based on an essentially perturbative treatment of the sideband transition. To validate this perturbative picture, the sideband transition has to be weaker than all the other relevant transitions, which severely limits the cooling rate. Until recently, ion cooling in the strong sideband coupling regime is considered, where it is shown that fast cooling could indeed be achieved in a dark-state cooling scheme with vanishing carrier transition~\cite{ZhangGuo2021}.

% The perturbative treatment of sideband cooling and dark-state cooling schemes often allow to derive analytical expressions for the cooling rate and the steady state average phonon occupation, such that the contribution of each parameter becomes transparent. However, such a perturbative treatment also limits the understanding of possible cooling mechanisms where the red sideband transition strength is comparable to other processes, therefore the possibility of fast cooling.

Another approach for ground state cooling is to think of cooling as an optimization problem whose target is to obtain the lowest possible average phonon occupation $\nave$ after applying certain control lasers, that is, employing the quantum control technique. Such an approach has been taken for trapped ion~\cite{MachnesRetzker2010}, and for mechanical resonators~\cite{WangJacobs2011,MachnesRetzker2012,LiuWong2013}. In particular, Refs.~\cite{MachnesRetzker2010,WangJacobs2011,MachnesRetzker2012} use complicated laser sequences optimized by quantum control such that the underlying ion or resonator could be cooled down close to the ground state in several periods of the trap frequency $\nu$ (super fast cooling), which could even be shorter than the dissipation time and the dissipation could then be neglected. Ref.~\cite{LiuWong2013} employs dynamical control of cavity dissipation technique in which the cavity dissipation is periodically enhanced, as a result ground state cooling could be achieved within several tens of periods of the trap frequency. However, the pulsed control laser sequences or periodically enhanced cavity dissipation add difficulty to realistic implementations. Till now, such schemes have not been demonstrated in ion cooling experiments yet to the best of our knowledge.

% Bulit on top of the traditional sideband and dark-state cooling
To harness the advantages from both the classical cooling schemes and the optimal quantum control technique, we propose a cooling strategy to enhance the classical cooling schemes with quantum control. Concretely, in our approach we use the same laser configuration (static lasers) as in the classical sideband or dark-state cooling schemes, but use optimal quantum control to search for the optimal cooling conditions. We will show that our approach could increase the cooling rate compared to the sideband cooling and dark-state cooling schemes but at the same time retain a low average phonon occupation. Compared to Ref.~\cite{ZhangGuo2021}, this approach allows a more systematic and accurate procedure to determine the optimal parameter settings. Additionally, most theoretical proposals for ion cooling only work in the Lamb-Dicke regime with $\eta \ll 1$, beyond this regime (for example in Penning trap the trap frequency is small, leading to a large Lamb-Dicke parameter) one often has to resort to numerical methods~\cite{RoghaniHelm2008,JoshiThompson2019}. Our method does not depend on the series expansion with $\eta$, which could provide a guidance for ion cooling experiments beyond the Lamb-Dicke regime.

This paper is organized as follows. In Sec.~\ref{sec:method}, we describe our approach to enhance the classical cooling schemes with optimal quantum control. In Sec.~\ref{sec:sbc}, we demonstrate our approach by optimizing the sideband cooling schemes including both the running wave sideband cooling and the standing wave sideband cooling, showing that the cooling rates of those schemes can be increased while still keeping $\nave$ close to the theoretical minimum. In Sec.~\ref{sec:eitc}, we further optimize the standard three-level electromagnetically induced transparency (EIT) cooling scheme as well as a realistic four-level EIT cooling scheme from the $^{40}$Ca$^{+}$ ion experiment. We conclude in Sec.~\ref{sec:summary}.

\section{Quantum control enhanced cooling scheme}\label{sec:method}
The dynamics of ion cooling in the Lamb-Dicke regime is often described by the Lindblad master equation~\cite{Lindblad1976,GoriniSudarshan1976}
\begin{align}\label{eq:lindblad}
\frac{d}{dt}\rhoop = \Lop(\rhoop) = -\im [\Hop, \rhoop] + \Dop(\rhoop),
\end{align}
where $\rhoop$ is the density operator of the system, $\Lop$ denotes the Lindblad operator (Lindbladian), $\Hop$ is the Hamiltonian and $\Dop$ is the dissipator in Lindblad form. In case $\Lop$ is time-independent, Eq.(\ref{eq:lindblad}) can be simply solved as 
\begin{align}\label{eq:linsol}
\rhoop(t) = e^{\Lop t} \rhoop_0 
\end{align}
with $\rhoop_0 = \rhoop(0)$ the initial state. Here we have assumed to squash the density operator $\rhoop$ as a vector and rewrite the Lindblad operator as a matrix accordingly~\cite{LandiSchaller2021}.

The elementary procedures of our approach can be summarized as follows. In sideband cooling or dark-state cooling schemes, the Lindbladian $\Lop$ is often parameterized by a few laser-related parameters such as the detuning and the Rabi frequency. We denote such a parametric Lindbladian as $\Lop(\paras)$ with $\paras$ to be a list of tunable parameters. The goal is to minimize the average phonon occupation of the quantum state after applying $\Lop(\paras)$ for a certain time interval $T$, for which the loss function can be naturally defined as
\begin{align}\label{eq:loss}
\loss_T(\paras) = \nave_T(\paras) = \trace(\nop \rhoop(T)),
\end{align}
with $\nop$ the phonon number operator. In practice it would often be more efficient if a gradient-based optimizer is used to minimize the loss function. The exact gradient of Eq.(\ref{eq:loss}) can be evaluated using an automatic differentiation package which supports complex numbers~\cite{GuoPoletti2021} and matrix functions such as matrix exponentiation. In this work we use the Zygote automatic differentiation framework implemented in Julia language. Zygote has a transparent support for both functionalities~\cite{Zygote2019}, with which we can simply write down the loss function and then the gradient could be obtained with almost no additional effort. The source code of all the simulations done in this work could be found at~\cite{sourcecode}. We will use the L-BFGS algorithm~\cite{LBFGS} as our optimization solver throughout this work.

To this end we note that our approach can be easily generalized to the case of pulsed lasers. For example, we could have a sequence of parametric Lindbladians $\{\Lop(\paras_1), \Lop(\paras_2), \dots, \Lop(\paras_n) \}$, applied onto the quantum state at time steps $\vec{T} = \{T_1, T_2, \dots, T_n\}$, the loss function in Eq.(\ref{eq:loss}) can then be straightforwardly generalized to
\begin{align}\label{eq:gloss}
\loss(\paras_1, \dots, \paras_n, \vec{T}) = \trace(\nop e^{\Lop(\paras_n)T_n} \dots e^{\Lop(\paras_1)T_1} \rhoop_0),
\end{align}
where $\vec{T}$ is explicitly placed as an input to the generalized loss function to indicate that they could also be treated as tunable parameters. A loss function in the form of Eq.(\ref{eq:gloss}) has been extensively considered in the context of optimal quantum control, see for example~\cite{2001Time,SchirmerFouquieres2011,2011Comparing,2011Second,Floether_2012,2014Evolutionary,2014Searching,JensenSherson2020,SchaferLorch2020,HernandezBrumer2021}. Here we note that automatic differentiation allows to directly compute the gradient of Eq.(\ref{eq:gloss}) as well, without resorting to more specific methods. In the context of this work, we will limit ourself to the loss function defined in Eq.(\ref{eq:loss}).

\section{Sideband cooling enhanced by quantum control}\label{sec:sbc}

Now we first exemplify our approach with running wave sideband cooling (RWSC)~\cite{CiracZoller1992}. In the ideal setup of RWSC, an ion of mass $m$ is trapped with trap frequency $\nu$ which contains two relevant internal states, a metastable ground state $\vert g\rangle$ with energy $\omega_g$ together with an excited state $\vert e\rangle$ with energy $\omega_e$ and \gcc{spontaneous decay rate $\gamma$ from $\vert e\rangle$ to $\vert g\rangle$}. The ion is coupled to a running wave laser with frequency $\omega_L$, wave number $k$ and Rabi frequency $\Omega$. The equation of motion is described by Eq.(\ref{eq:lindblad}) with the Hamiltonian $\Hop_{{\rm RW}}$
\begin{align}\label{eq:rwh}
\Hop_{{\rm RW}}(\Delta, \Omega) =& -\Delta \vert e\rangle\langle e\vert + \nu \adop\aop \nonumber \\
&+ \frac{\Omega}{2}\left( \vert e\rangle\langle g\vert e^{-\im k\xop} + \vert g\rangle\langle e\vert e^{\im k\xop} \right),
\end{align}
where $\Delta = \omega_L - (\omega_e - \omega_g)$ is the detuning, $\adop$ and $\aop$ creates and annihilates a phonon, $\xop = \frac{1}{\sqrt{2m\nu}}(\adop + \aop) $ is the positional operator. With the Lamb-Dicke parameter defined as $\eta=k/\sqrt{2m\nu}$, we get $k\xop = \eta(\adop + \aop)$. Here we have explicitly written $\Hop_{{\rm RW}}(\Delta, \Omega)$ to indicate that $\Delta$ and $\Omega$ are two tunable parameters in usual experimental setups. The dissipator $ \Dop_{{\rm RW}}$ takes the form
\begin{align}\label{eq:rwd}
\Dop_{\textrm{RW}}(\rhoop)= & \frac{\gamma}{2}\int_{-1}^{1}d\left(\cos(\theta)\right)\left(\frac{3}{4}\left(1+\cos^{2}(\theta)\right)\right)\vert g\rangle\langle e\vert\nonumber \\
 & e^{\im k\xop\cos(\theta)}\rhoop e^{-\im k\xop\cos(\theta)}\vert e\rangle\langle g\vert-\frac{\gamma}{2}\{\vert e\rangle\langle e\vert,\rhoop\}.
\end{align}
$\Dop_{{\rm RW}}$ is not tunable in general. The classical approach to find the optimal cooling condition is to perturbatively expand Eqs.(\ref{eq:rwh}, \ref{eq:rwd}) in $\eta$ and then truncate $\Hop_{{\rm RW}}$ and $\Dop_{{\rm RW}}$ to the first order of $\eta$. Under this approximation, the analytic expression for the steady state average phonon occupation $\nss_{{\rm RW}}$ can be obtained as~\cite{CiracZoller1992}
\begin{align}\label{eq:nssrw}
\nss_{{\rm RW}} = \frac{A_+^{{\rm RW}}}{A_-^{{\rm RW}} - A_+^{{\rm RW}}},
\end{align}
with $A^{{\rm RW}}_{\pm} = \eta^2\left(\frac{\Omega}{2} \right)^2\left(\frac{\gamma}{(\Delta\mp\nu)^2 + \gamma^2 / 4} + \frac{0.4\gamma}{\Delta^2 + \gamma^2 / 4} \right)$. Eq.(\ref{eq:nssrw}) is only valid under the conditions that $\gamma, \Omega \ll \nu$ (resolved sideband condition) and $\eta\Omega \ll \gamma$ (weak sideband coupling condition). The theoretical minimum of $\nss_{{\rm RW}}$ is reached at $\Delta = -\nu$, that is, when the red sideband resonance condition is satisfied.

\begin{figure}
\includegraphics[width=\columnwidth]{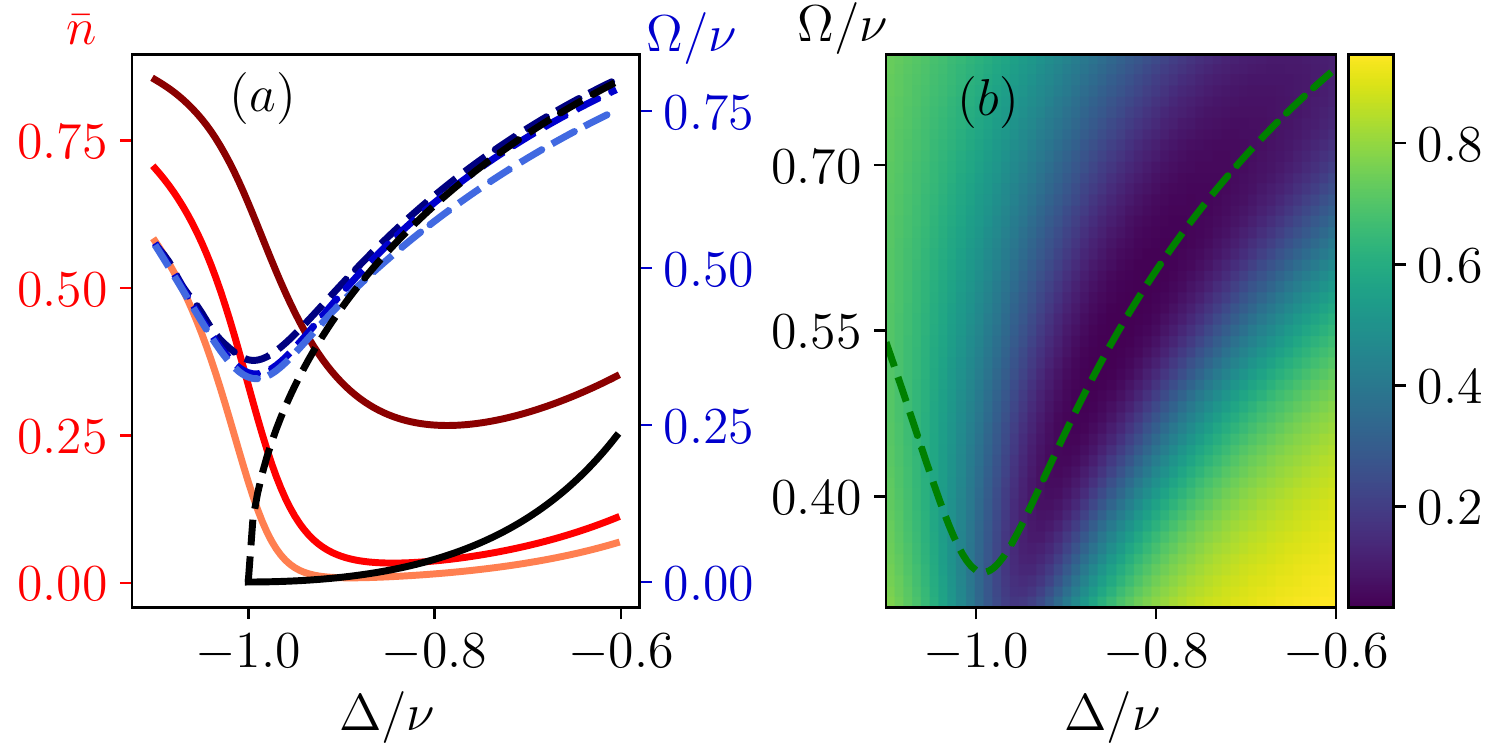}
\caption{Quantum control enhanced running wave sideband cooling. (a) The blue dashed lines from top down and from darker to lighter shows the optimal values of $\Omega$ as functions of $\Delta$ for $T\nu=100, 250, 400$ respectively, while the solid lines from top down and from darker to lighter show the corresponding values of $\nave_T$. For example, at $T\nu=250$ and $\Delta=-0.8\nu$, the optimal value of $\Omega$ to minimize $\nave_T$ is $\Omega=0.057\nu$ from the middle blue dashed line, with the corresponding minimal value $\nave_T=0.015$ which can be read from the  middle red solid line. The black solid line shows $\nss_{{\rm RW}}$ as a function of $\Delta$ as predicted from Eq.(\ref{eq:nssrw}), while the black dashed line shows the shifted red sideband resonance condition as in Eq.(\ref{eq:srsr}). (b) $\nave_T$ as a function of $\Delta$ and $\Omega$ at $T\nu=250$. The green dashed line shows the optimal value of $\Omega$ as a function of $\Delta$ as in (a). The other parameters used are $\eta=0.1$, $\nu=1$, $\gamma=0.1\nu$. The initial state of the phonon is chosen as a thermal state with $\nave_0 = 1$ and \gcc{we use a truncation $d=10$ for the Fock state space of the phonon}.
}
\label{fig:fig1}
\end{figure}

Now we can directly substitute $\Lop_{{\rm RW}}(\Delta, \Omega)(\rhoop) = -\im[\Hop_{{\rm RW}}(\Delta, \Omega), \rhoop] + \Dop_{{\rm RW}}(\rhoop) $ into Eq.(\ref{eq:loss}) and then obtain the optimal values of $\Delta$ and $\Omega$ for each value of $T$ using an optimization solver. However to better visualize the effect of the numerical optimization, we first fix the value of $\Delta$ and then find the optimal value of $\Omega$ which minimizes $\nave_T$. The simulation results are shown in Fig.~\ref{fig:fig1}, where we have considered $T\nu=100, 250, 400$ respectively. For each value of $T$, we change $\Delta$ from $-1.1\nu$ to $-0.6\nu$, and for each value of $\Delta$ we find the optimal $\Omega$ by optimizing Eq.(\ref{eq:loss}) as a single-variate function of $\Omega$. The optimal values of $\Omega$ as functions of $\Delta$ are shown in dashed lines in Fig.~\ref{fig:fig1}(a) (from top down, the blue dashed lines correspond to $T\nu=100, 250, 400$ respectively). The corresponding optimal values of $\nave_T$ are shown in solid lines from top down. For reference, we also show the theoretical prediction from Eq.(\ref{eq:nssrw}) in black solid line. We can see that the optimal values of $\Delta$ shift to the right hand side of the red sideband resonant point for all the $T$s we have considered. 
% For a moderate value of $T\nu$ to be several hundreds of periods of the trap frequency, to reach a low $\nave_T$, the Rabi frequency has to be increased. 
From Fig.~\ref{fig:fig1}(a), we can see that $\nave_T$ reaches its minimum when $\Omega \approx 0.4\nu$ and $\eta\Omega / \gamma \approx 0.4$, where the validity of Eq.(\ref{eq:nssrw}) is no longer justified. Nevertheless for a moderate value of $T\nu=400$, we could already reach a minimum value with $\nave_T=0.0087$ at $\Delta=-0.891\nu$ and $\Omega=0.459\nu$, in comparison with the minimum value $\nss_{{\rm RW}}=0.0016$ predicted by Eq.(\ref{eq:nssrw}).
In fact, it often takes several thousands of periods of the trap frequency to reach the theoretical minimum of $\nss_{{\rm RW}}$ under the weak sideband coupling condition~\cite{ZhangGuo2021}. 
Therefore, optimal quantum control allows to achieve faster cooling in classical sideband cooling by one order of magnitude with only a small increase of the final average phonon occupation. Interestingly, in this case we could impose a modified red sideband resonance condition to understand the relation between the optimal values of $\Omega$ and $\Delta$. Diagonalizing the Hamiltonian for the internal two-level system, we get the energy difference $\sqrt{\Delta^2 + \Omega^2}$, then the shifted red sideband resonance condition reads (one can also see Ref.~\cite{MarzoliZoller1994})
\begin{align}\label{eq:srsr}
\sqrt{\Delta^2 + \Omega^2} = \nu,
\end{align}
which is plotted in black dashed line in Fig.~\ref{fig:fig1}(a). We can see that it agrees well with the blue dashed lines when $\Delta > -0.9\nu$. To further visualize the effect of the numerical optimization, in Fig.~\ref{fig:fig1}(b), we directly plot $\nave_T$ for all the values of $\Delta/\nu$ in the range $[-1.1, -0.6]$ and $\Omega/\nu$ in the range $[0.3, 0.8]$ at $T\nu=250$. The optimized value of $\Omega$ as a function of $\Delta$ is also shown in green dashed line, and we can see that it well captures the minimum values of $\nave_T$.

\gcc{It would also be instructive to consider the effects of thermal states with higher initial average occupation $\nave_0$. Here we consider a thermal state with $\nave_0=3$ and use a truncation $d=30$ as an example. Concretely, we optimize the set of parameters with $\nave_0=3$ and a total evolution time $T\nu=636$, the latter is chosen such that it will result in the same $\nave_T=0.0087$ if subjecting to the same exponential decay rate of $0.011/\nu$ as fitted from the case of $\nave_0=1$ and $T\nu=400$. Interestingly, in this case we get $\Delta=-0.895\nu$ and $\Omega=0.464\nu$, which is almost the same as the case of $\nave_0=1$ and $T\nu=400$, while the cooling rate is $0.0073/\nu$, which is slightly lower. In practice, it would be interesting to consider thermal states with $\nave_0 \geq 10$ which is often the starting point for sideband cooling experiments~\cite{roosthesis}. However, to faithfully represent a thermal state with $\nave_0 \approx 10$, one has to reserve a large number of Fock states (often more than $100$), which is numerically demanding with our current approach since we explicitly store the Lindbladian as a dense matrix. To scale up the simulation one could use either a larger workstation or explore the sparsity of the Lindbladian, but in the latter case the back propagation of certain matrix functions has to be explicitly defined and we will leave it for future investigation.}

\begin{figure}
\includegraphics[width=\columnwidth]{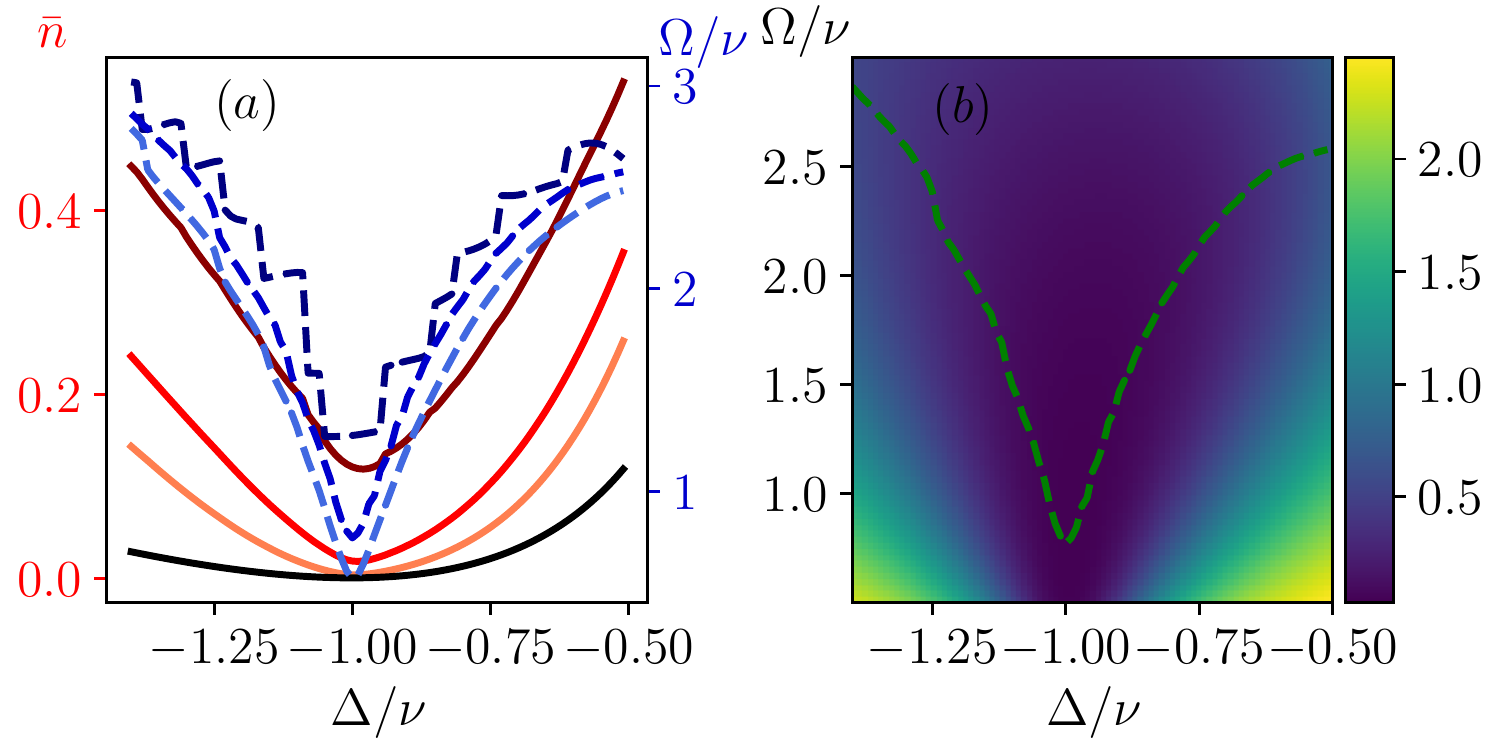}
\caption{Quantum control enhanced standing wave sideband cooling. (a)The blue lines from top down and from darker to lighter shows the optimal values of $\Omega$ as functions of $\Delta$ for $T\nu=80, 160, 240$ respectively, while the solid red lines from top down and from darker to lighter show the corresponding values of $\nave_T$. The black solid line shows $\nss_{{\rm SW}}$ as a function of $\Delta$ as predicted from Eq.(\ref{eq:nsssw}). (b) $\nave_T$ as a function of $\Delta$ and $\Omega$ at $T\nu=160$. The green dashed line shows the optimal value of $\Omega$ as a function of $\Delta$ as in (a). The other parameters used are $\eta=0.08$, $\gamma=0.1\nu$. The initial state of the phonon is chosen as a thermal state with $\nave_0 = 1$ and \gcc{we use a truncation $d=10$ for the Fock state space of the phonon}.
}
\label{fig:fig2}
\end{figure}

In the next we consider the case of standing wave sideband cooling (SWSC). The setup is similar to the case of RWSC with the only difference that a standing wave laser is used and the ion is placed at the node of it. The Hamiltonian $\Hop_{{\rm SW}}$ can be written as
\begin{align}\label{eq:swh}
\Hop_{{\rm SW}}(\Delta, \Omega) =& -\Delta \vert e\rangle\langle e\vert + \nu \adop\aop \nonumber \\
&+ \frac{\Omega}{2}\left( \vert e\rangle\langle g\vert + \vert g\rangle\langle e\vert \right) \sin(k\xop) .
\end{align}
The dissipator is the same with the case of RWSC, that is, $\Dop_{{\rm SW}} = \Dop_{{\rm RW}}$. In this case the steady state average phonon occupation $\nss_{{\rm SW}}$ is predicted as~\cite{CiracZoller1992}
\begin{align}\label{eq:nsssw}
\nss_{{\rm SW}} = \frac{A_+^{{\rm SW}}}{A_-^{{\rm SW}} - A_+^{{\rm SW}}},
\end{align}
with $A^{{\rm SW}}_{\pm} = \eta^2\left(\frac{\Omega}{2} \right)^2 \frac{\gamma}{(\Delta\mp\nu)^2 + \gamma^2 /4} $.

Now we perform a similar numerical optimization procedure to Fig.~\ref{fig:fig1} with $\Lop_{{\rm SW}}(\Delta, \Omega)(\rhoop) = -\im[\Hop_{{\rm SW}}(\Delta, \Omega), \rhoop]+ \Dop_{{\rm SW}}(\rhoop)$ and the results are shown in Fig.~\ref{fig:fig2} with $T\nu=80, 160, 240$ respectively. For each value of $T$, we change $\Delta$ from $-1.4\nu$ to $-0.5\nu$, and for each value of $\Delta$ we find the optimal value of $\Omega$. The optimal values of $\Omega$ as functions of $\Delta$ are shown in dashed lines in Fig.~\ref{fig:fig2}(a) (from top down, the blue dashed lines correspond to $T\nu=80, 160, 240$ respectively). The corresponding optimal values of $\nave_T$ are shown
in red solid lines. In this case we can see that $\nave_T$ always reaches its minimum at the red sideband resonance point $\Delta = -\nu$, in comparison with RWSC. We can also see that $\Omega \approx 0.575\nu$ and $\eta\Omega/\gamma \approx 0.46$ when $\nave_T$ reaches its minimum at $T\nu=240$, which is clearly in the strong sideband coupling regime. Moreover, at $T\nu=240$, we could already reach an optimal value of $\nave_T=0.0039$, which is close to the theoretical minimum $\nss_{{\rm SW}} = 0.0006$ (The theoretical minimum itself is derived under several approximations and may not be reachable by solving the exact Lindblad master equation). These results are in correspondence with the results in~\cite{ZhangGuo2021}. In Fig.~\ref{fig:fig2}(b), we plot all the values of $\nave_T$ at $T\nu=160$ for $\Delta/\nu$ in range $[-1.4, -0.5]$ and $\Omega/\nu$ in range $[0.5, 3]$. The optimal value of $\Omega$ as a function of $\Delta$ is also shown in blue dashed line for reference.

% In Fig.~\ref{fig:fig2}(a), we plot $\nave_T$ as a function of $\Delta$ (solid lines) for $T=80, 160, 240$ respectively, with the optimal values of $\Omega$ shown in dashed lines. In comparison with Fig.~\ref{fig:fig1}(a), we can see from Fig.~\ref{fig:fig2}(a) that $\nave_T$ reaches its minimum at $\Delta=-\nu$ for all the values of $T$s considered, that is, the red sideband resonance condition is exact satisfied by the optimal values. The black solid line shows the steady state $\nss$ predicted in the perturbative regime as

% We can see that for $T=240$, we could already reach an optimal value of $\nave_T=XXX$, which is very close to the theoretical prediction $\nss = XXX$. We can also see that the optimial values of $\Omega$ in general does not satisfy the weak sideband condition $\eta\Omega \ll \gamma$, as pointed out in Ref.~\cite{ZhangGuo2021}.

\section{EIT cooling enhanced by quantum control}\label{sec:eitc}

\begin{figure}
\includegraphics[width=\columnwidth]{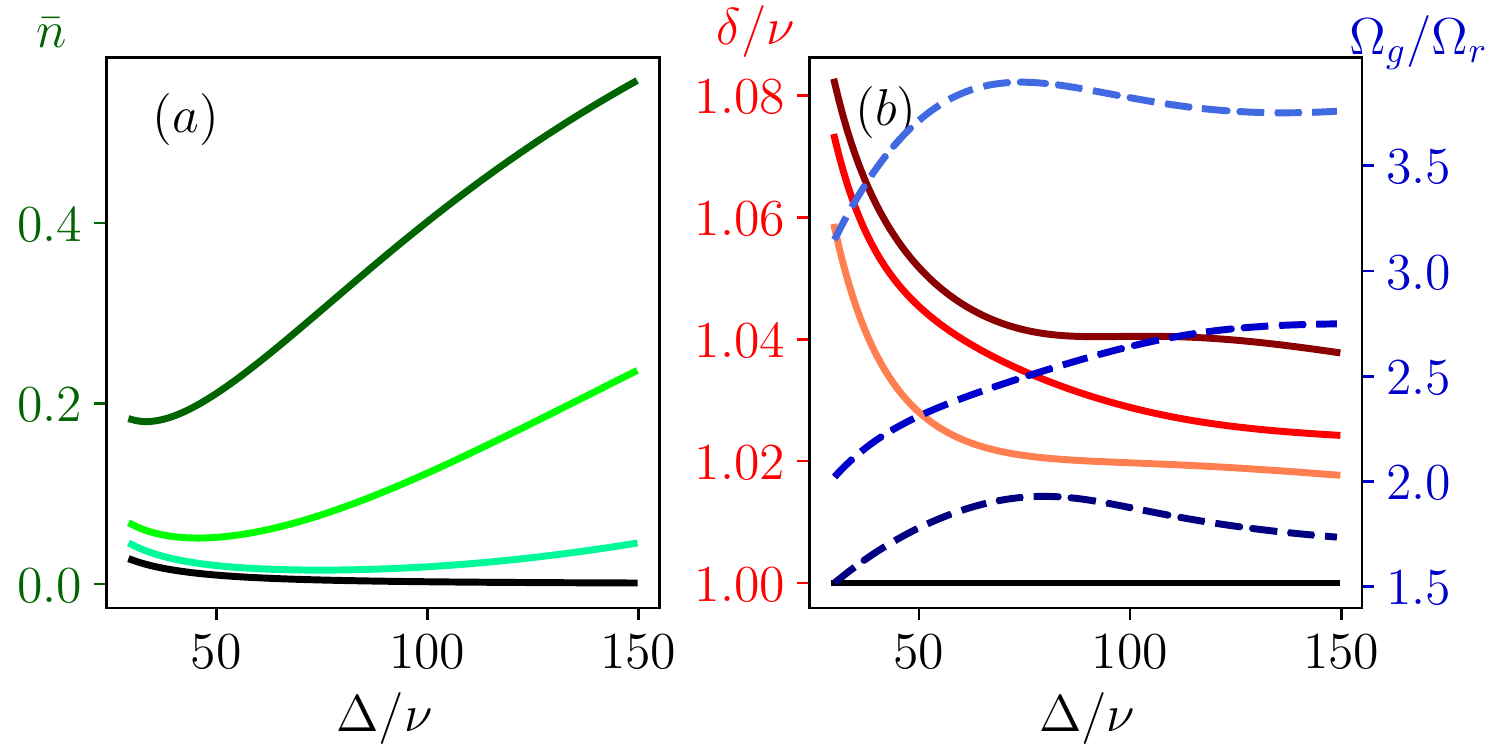}
\caption{Quantum control enhanced standard three-level EIT cooling. (a) The green solid lines from top down and from darker to lighter shows the minimum values of $\nave_T$ as a function of $\Delta$ for $T\nu=50, 100, 200$ respectively. The black solid line shows $\nss_{{\rm EIT}}$ as a function of $\Delta$ as predicted from Eq.(\ref{eq:nsseit}) under the condition in Eq.(\ref{eq:eitcond}). (b) The red solid lines from top down and from darker to lighter shows the AC Stark shift $\delta$ as a function of $\Delta$, computed at the corresponding optimal values of $\Omega_g$ and $\Omega_r$, while the blue dashed lines from down to top and from lighter to darker plot the ration $\Omega_g / \Omega_r$. The black solid line shows the theoretical optimal value of $\delta$, namely $\delta /\nu=1$. The other parameters used are $\eta_g = 0.15$, $\eta_r = -0.15$, $\gamma_g = 20/3 \nu$, $\gamma_r = 40/3 \nu$. The initial state of the phonon is chosen as a thermal state with $\nave_0 = 1$ and \gcc{we use a truncation $d=10$ for the Fock state space of the phonon}.
}
\label{fig:fig3}
\end{figure}

Now we apply our strategy to more elaborated dark-state cooling schemes. To start we consider the standard EIT cooling scheme which can be implemented using a $\Lambda-$type three-level
structure with an excited state $\left|e\right\rangle $ of \gcc{spontaneous decay rate}
$\gamma$, and two metastable ground states $\vert g\rangle$ and
$\vert r\rangle$. Two lasers are used to couple $\vert g\rangle\leftrightarrow\vert e\rangle$
and $\vert r\rangle\leftrightarrow\vert e\rangle$, with frequencies
$\omega_{g}$ and $\omega_{r}$, wave numbers $k_{g}$ and $k_{r}$,
Rabi frequencies $\Omega_{g}$ and $\Omega_{r}$ respectively. The
Hamiltonian $\Hop_{{\rm EIT}}$ can be written as 
\begin{align}
&\Hop_{{\rm EIT}}(\Delta, \Omega_g, \Omega_r) \nonumber \\
=& -\Delta\vert e\rangle\langle e\vert+\nu\adop\aop \nonumber \\
 & + \frac{\Omega_{g}}{2}\left(\vert g\rangle\langle e\vert e^{-\im k_{g}\xop}+\vert e\rangle\langle g\vert e^{\im k_{g}\xop}\right) \nonumber \\ 
 & + \frac{\Omega_{r}}{2}\left(\vert r\rangle\langle e\vert e^{\im k_{r}\xop}+\vert e\rangle\langle r\vert e^{-\im k_{r}\xop}\right),
\end{align}
with $\Delta$ the detuning for both lasers (one could refer to Ref.~\cite{MorigiKeitel2000,ZhangGuo2021} for example for the standard EIT level structure). Here we have assumed that both lasers are in parallel with the motional axis of the ion and that the two lasers propagate in opposite directions. We have also written $\Hop_{{\rm EIT}}(\Delta, \Omega_g, \Omega_r)$ to indicate that the parameters $\Delta$, $\Omega_g$, $\Omega_r$ are easily tunable in usual experimental settings.
% $\eta_{g}=k_{g}/\sqrt{2m\nu}$ and $\eta_{r}=k_{r}/\sqrt{2m\nu}$
The dissipator $\Dop_{\textrm{EIT}}$ can be written as
\begin{align}
 & \Dop_{\textrm{EIT}}(\rhoop)\nonumber \\
= & \sum_{j=g,r} \frac{\gamma_{j}}{2} \int_{-1}^{1}d\left(\cos(\theta)\right)\left(\frac{3}{4}\left(1+\cos^{2}(\theta)\right)\right)\vert j\rangle\langle e\vert\nonumber \\
 & e^{\im k_{j}\xop\cos(\theta)}\rhoop e^{-\im k_{j}\xop\cos(\theta)}\vert e\rangle\langle j\vert-\frac{\gamma_{j}}{2}\{\vert e\rangle\langle e\vert,\rhoop\}.
\end{align}
with $\gamma_{g}$ and $\gamma_{r}$ the decay rates from $\vert e\rangle$
to $\vert g\rangle$ and to $\vert r\rangle$ respectively. The \gcc{spontaneous decay rate} $\gamma$ of the excited state $\vert e\rangle$ satisfies $\gamma = \gamma_g + \gamma_r$. The Lamb-Dicke parameters corresponding to the two lasers are defined as $\eta_{g}=k_{g}/\sqrt{2m\nu}$ and $\eta_{r}=k_{r}/\sqrt{2m\nu}$. The steady state average phonon occupation for EIT cooling has been derived as~\cite{MorigiKeitel2000,Morigi2003}
\begin{align}\label{eq:nsseit}
\nss_{{\rm EIT}} = \frac{A_+^{{\rm EIT}}}{A_-^{{\rm EIT}} - A_+^{{\rm EIT}}},
\end{align}
with $A^{{\rm EIT}}_{\pm} = \frac{\Omega_g^2}{\gamma} \frac{\gamma^2\nu^2}{\gamma^2\nu^2 + 4\left(\left(\Omega_g^2 + \Omega_r^2\right)/4 - \nu(\nu \mp \Delta) \right)^2} $, where $\Omega_g \ll \Omega_r$ is assumed. The optimal cooling condition to minimize Eq.(\ref{eq:nsseit}) is obtained by requiring the AC Stark shift $\delta$ of internal bright state, defined as 
\begin{align}\label{eq:delta}
\delta = \frac{1}{2}\left(-\Delta + \sqrt{\Omega_g^2 + \Omega_r^2 + \Delta^2} \right),
\end{align}
to match the trap frequency ~\cite{MorigiKeitel2000,Morigi2003,ZhangGuo2021}
\begin{align}\label{eq:eitcond}
\delta = \nu .
\end{align}

Now similar to the case of sideband cooling, for each $T$, we first fix the value of $\Delta$ in $\Lop_{{\rm EIT}}(\Delta, \Omega_g, \Omega_r)(\rhoop) = -\im[\Hop_{{\rm EIT}}(\Delta, \Omega_g, \Omega_r), \rhoop] + \Dop_{{\rm EIT}}(\rhoop)$ and then find the optimal values of $\Omega_g$ and $\Omega_r$ as functions of $\Delta$. The simulation results are shown in Fig.~\ref{fig:fig3}, where we have considered $T\nu = 50, 100, 200$ respectively. For each value of $T$, we change $\Delta$ from $30\nu$ to $150\nu$, and for each value of $\Delta$ we compute the optimal values of $\Omega_g$ and $\Omega_r$ by optimizing Eq.(\ref{eq:loss}). The optimal values of $\nave_T$ as functions of $\Delta$ are shown in green solid lines from top down in Fig.~\ref{fig:fig3}(a), while the black solid line represents the theoretical minima by substituting Eq.(\ref{eq:eitcond}) into Eq.(\ref{eq:nsseit}).
We can see that for $T\nu = 200$ we could already reach $\nave_T = 0.0154$ which is close to the theoretical minimum $\nss_{{\rm EIT}}=0.0045$. 
In Fig.~\ref{fig:fig3}(b) we plot the AC Stark shift $\delta$ as a function of $\Delta$, which is shown by the red solid lines from top down and from darker to lighter, corresponding to $T\nu=50,100,200$ respectively. We also plot the ratio $\Omega_g / \Omega_r$ as a function of $\Delta$ in blue dashed lines from bottom up and from lighter to darker correspondingly. The theoretical optimal EIT cooling condition in Eq.(\ref{eq:eitcond}) is shown in black solid line in Fig.~\ref{fig:fig3}(b) for reference. From the solid lines we can see that the the optimal values of $\Delta$, $\Omega_g$ and $\Omega_r$ still approximately satisfy Eq.(\ref{eq:eitcond}). However, from the dashed lines we can see that $\Omega_g > \Omega_r$ while in standard derivation of Eq.(\ref{eq:nsseit}) the opposite is often assumed~\cite{MorigiKeitel2000}. This is because we have chosen $\gamma_g < \gamma_r$, thus to achieve fast cooling, it would be better if the dark state is mostly in the state $\vert r\rangle$, as a result the role of $\vert g\rangle$ and $\vert r\rangle$ has been interchanged.

% In standard EIT derivation $\delta$ is chosen such that $\delta = \nu$.

\begin{figure}
\includegraphics[width=\columnwidth]{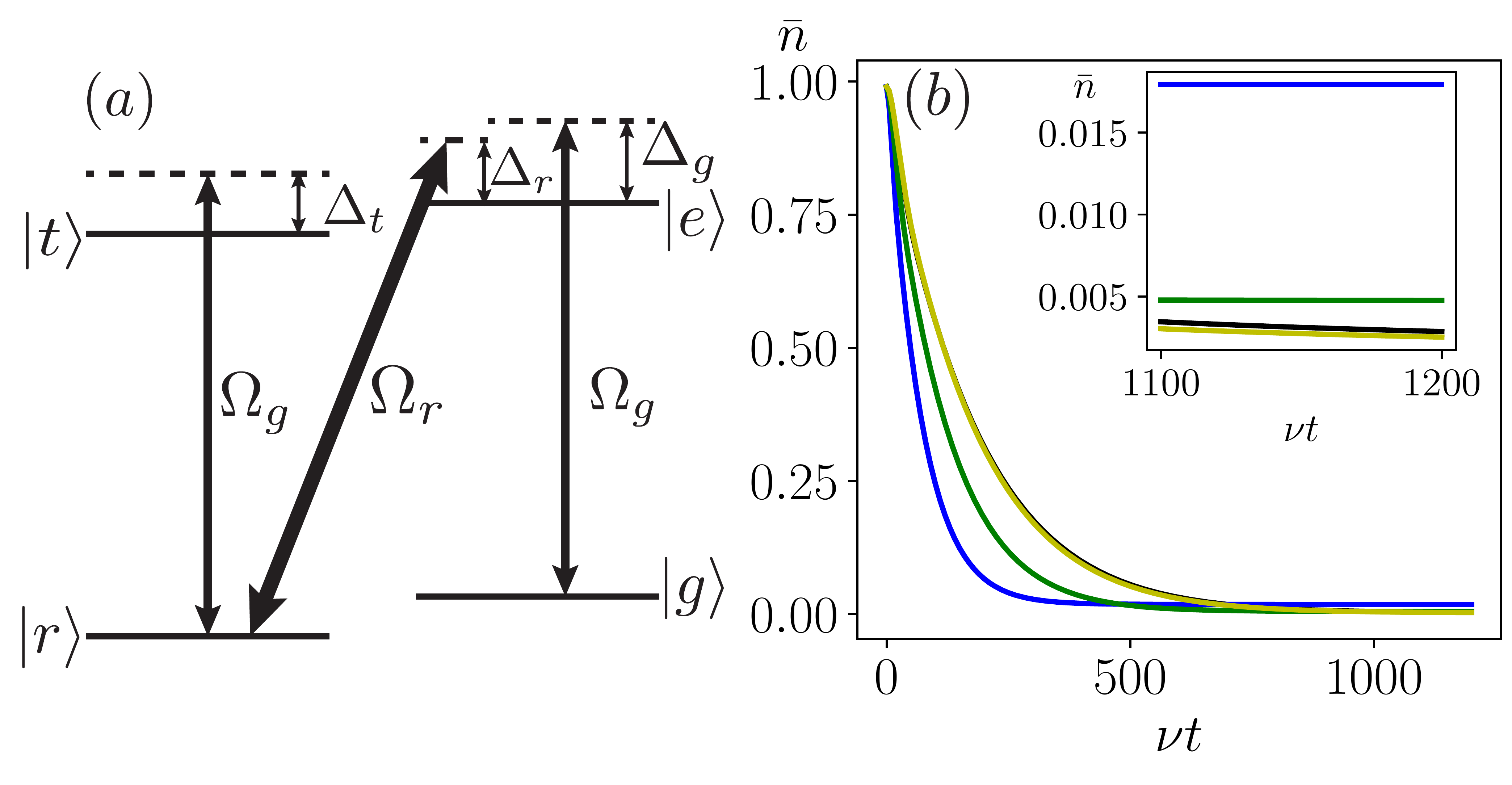}
\caption{Quantum control enhanced EIT cooling in a realistic four-level structure. (a) Level structure for the $^{40}$Ca$^{+}$ ion used for EIT cooling. A $\pi$-polarized laser couples the two transitions $\vert g\rangle \leftrightarrow \vert e\rangle$ and $\vert r\rangle \leftrightarrow \vert t\rangle$ with detunings $\Delta_g$ and $\Delta_t$, and Rabi frequency $\Omega_g$. A $\sigma^+$ polarized laser couples the transition $\vert r\rangle \leftrightarrow \vert e\rangle$ with detuning $\Delta_r$ and Rabi frequency $\Omega_r$. (b) Time evolution of $\nave(t)$ with the optimal set of parameters ($\Delta_g$, $\Delta_r$, $\Omega_g$, $\Omega_r$) for $T\nu=300$ (blue line), $T\nu=700$ (green line) and $T\nu=1200$ (yellow line) respectively. The black line represents the time evolution with the optimal set of parameters computed by assuming $\Omega_g \ll \Omega_r$ and neglecting the energy level $\vert t\rangle$. The inset shows the time evolution starting from $\nu t = 1100$. The other parameters used are $\nu=1.3\times 2\pi $ MHz, $\Omega_t = \Delta_g + 8\nu$. The initial state of the phonon is chosen as a thermal state with $\nave_0 = 1$ and \gcc{we use a truncation $d=10$ for the Fock state space of the phonon}.
}
\label{fig:fig4}
\end{figure}

The powerfulness of our approach is best demonstrated in a realistic experimental setting in which there are usually unwanted transitions except those required by an ideal setup. Here we take the case of EIT cooling implemented with a $^{40}$Ca$^{+}$ ion as an example. In this case EIT cooling is often implemented on the $^{1/2}S\rightarrow{}^{1/2}P$ transition. The two Zeeman sublevels $^{1/2}S$ and $^{1/2}P$ form
a four-level system. A $\sigma^+$ polarized laser couples $\vert S, -\rangle \leftrightarrow \vert P, +\rangle$ with detuning $\Delta_r$ and Rabi frequency $\Omega_r$. A $\pi$ polarized laser couples the transitions $\vert S, -\rangle \leftrightarrow \vert P, -\rangle$ and $\vert S, +\rangle \leftrightarrow \vert P, +\rangle$ with Rabi frequency $\Omega_g$ and with detunings $\Delta_t$ and $\Delta_g$ respectively. In the next we denote $\vert g\rangle = \vert S, +\rangle$, $\vert r\rangle = \vert S, -\rangle$, $\vert e\rangle = \vert P, +\rangle$ and $\vert t\rangle = \vert P, -\rangle$ for briefness. The level diagram is shown in Fig.~\ref{fig:fig4}(a). The Hamiltonian $\Hop_{{\rm EIT4}}$ can be written as
\begin{align}\label{eq:reith}
&\Hop_{{\rm EIT4}}(\Delta_{g}, \Delta_r, \Omega_{g}, \Omega_r) \nonumber \\
=& \nu \adop\aop - \Delta_g \vert e\rangle\langle e\vert - (\Delta_g -\Delta_r) \vert r\rangle\langle r\vert \nonumber \\
&+  (-\Delta_g + \Delta_r - \Delta_t) \vert t\rangle \langle t \vert \nonumber \\
& + \frac{\Omega_g}{2}\left(\vert g\rangle \langle e\vert e^{\im \frac{\sqrt{2}}{2} \eta (\adop + \aop) } + \hc \right) \nonumber \\
& + \frac{\Omega_r}{2}\left(\vert r\rangle \langle e\vert e^{-\im \frac{\sqrt{2}}{2} \eta (\adop + \aop) } + \hc \right) \nonumber \\
& + \frac{\Omega_g}{2}\left(\vert r\rangle \langle t\vert e^{\im \frac{\sqrt{2}}{2} \eta (\adop + \aop) } + \hc \right),
\end{align}
where we have explicitly written $\Hop_{{\rm EIT4}}(\Delta_{g}, \Delta_r, \Omega_{g}, \Omega_r)$ to indicate that $\Delta_{g}, \Delta_r, \Omega_{g}, \Omega_r$ are the four independent tunable parameters for the setup (the difference between $\Delta_t$ and $\Delta_g$ arises from the different energy splitting of the $S_{1/2}$ and $P_{1/2}$ Zeeman sublevels in a magnetic field, and we have chosen $\Delta_t = \Delta_g + 8 \nu$). Here we have assumed that the angle between the motional axis and the magnetic field is $\pi / 4$, and that the $\sigma^+$-polarized laser propagates along the magnetic field while the $\pi$-polarized laser is orthogonal to it, such that the effective Lamb-Dicke parameter is $\sqrt{2}\eta$. The dissipator $\Dop_{{\rm EIT4}}$ can be written as
\begin{align}\label{eq:reitd}
\Dop_{{\rm EIT4}}(\rhoop) =& \frac{\gamma_g}{2}\left(2\vert g\rangle\langle e\vert \tilde{\rho}_1 \vert e\rangle\langle g\vert -\{\vert e\rangle \langle e\vert, \rhoop \} \right) \nonumber \\
&+ \frac{\gamma_r}{2}\left(2\vert r\rangle\langle e\vert \tilde{\rho}_2 \vert e\rangle\langle r\vert -\{\vert e\rangle \langle e\vert, \rhoop \} \right) \nonumber \\
& + \frac{\gamma_g}{2}\left(2\vert g\rangle\langle t\vert \tilde{\rho}_2 \vert t\rangle\langle g\vert -\{\vert t\rangle \langle t\vert, \rhoop \} \right) \nonumber \\
& + \frac{\gamma_r}{2}\left(2\vert r\rangle\langle t\vert \tilde{\rho}_1 \vert t\rangle\langle r\vert -\{\vert t\rangle \langle t\vert, \rhoop \} \right),
\end{align}
with
\begin{align}
\tilde{\rho}_1 &= \int_{-1}^1 dx \left( \frac{3}{4}(1 + x^2)e^{-\im \eta (\aop+\adop)x} \rhoop e^{\im \eta (\aop+\adop)x} \right); \\
\tilde{\rho}_2 &= \int_{-1}^1 dx \left( \frac{3}{2}(1 - x^2)e^{-\im \eta (\aop+\adop)x} \rhoop e^{\im \eta (\aop+\adop)x} \right).
\end{align}
Here \gcc{$\gamma_g=2\pi \times 20/3$ MHz} is the decay rate from $\vert e\rangle$ to $\vert g\rangle$ and from $\vert t\rangle$ to $\vert r\rangle$, and \gcc{$\gamma_r=2\pi \times 40/3$ MHz} is the decay rate from $\vert e\rangle$ to $\vert r\rangle$ and from $\vert t\rangle$ to $\vert g\rangle$, \gcc{the ratio between $\gamma_g$ and $\gamma_r$ is fixed by the Clebsch-Gordan cofficients}. The slight difference between the expressions of $\tilde{\rho}_1$ and $\tilde{\rho}_2$ is due to the different emission dipole patterns~\cite{ReibBlatt1996}.

\begin{table}
\caption{Optimal EIT cooling conditions in a realistic four-level structure. The first column is the period fixed for optimal quantum control. The second to the fifth columns show the optimal values of the parameters $\Delta_g$, $\Omega_r$, $\Omega_g$ and $\Omega_r$ for the different values of $T$ respectively. The sixth column show the cooling rate from exponential fitting of the real time evolution, and the last column show the final average phonon occupation at $t = 1200/\nu$ by using the corresponding set of optimal parameters. The last row shows the corresponding set of optimal parameters based on the simplified three-level scheme with a fixed period $T\nu=1200$.}
\label{tab:fourleveloptimal}
\centering
\begin{tabular*}{\columnwidth}{c @{\extracolsep{\fill}} ccccccc}
\Xhline{1pt}
$T \nu $  & $\Omega_g / \nu$ & $\Omega_r / \nu$  & $\Delta_g / \nu $ & $\Delta_r / \nu $ & $W / \nu$ & $\nave_{1200} $ \\
\hline
300    & 3.7   & 17.1 &  65.9   & 65.7 & $14.6 \times 10^{-3}$ & $27.7 \times 10^{-3}$  \\
700    & 2.7   & 22.0 & 109.2   & 109.0 & $8.7 \times 10^{-3}$ & $6.4 \times 10^{-3}$ \\
1200    & 2.3   & 26.7 & 159.7   & 159.6 & $5.9 \times 10^{-3}$ & $2.5 \times 10^{-3}$  \\
\Xhline{0.5pt}
EIT3-1200    & 2.3   & 24.5 & 150.4   & 150.4 & $5.9 \times 10^{-3}$ & $2.9 \times 10^{-3}$  \\
\Xhline{1pt}
\end{tabular*}
\end{table}

The existence of the additional level $\vert t\rangle$ makes it extremely difficult to derive an analytic expression for $\nss$ as well as an optimal cooling condition similar to the standard three-level EIT, especially in the strong sideband coupling regime. However, using quantum control we can compute the optimal cooling condition numerically with almost the same effort as before. Now we substitute $\Lop_{{\rm EIT4}}(\Delta_{g}, \Delta_r, \Omega_{g}, \Omega_r)(\rhoop) = -\im[\Hop_{{\rm EIT4}}(\Delta_{g}, \Delta_r, \Omega_{g}, \Omega_r), \rhoop] +\Dop_{{\rm EIT4}}(\rhoop)$ into Eq.(\ref{eq:loss}) and compute the optimal values of $\Delta_{g}, \Delta_r, \Omega_{g}, \Omega_r$ at $T\nu=300, 700, 1200$ respectively. We then substitute those optimal values into Eq.(\ref{eq:lindblad}) with Hamiltonian from Eq.(\ref{eq:reith}) and dissipator from Eq.(\ref{eq:reitd}), and evolve the initial state for a time $t=1200/\nu$. The simulation results are shown in Fig.~\ref{fig:fig4}(b), where the blue, green, yellow solid lines correspond to the evolution based on the optimal parameters found for $T\nu=300, 700, 1200$ respectively. In practice, the transition $\vert r\rangle \leftrightarrow \vert t\rangle$ is often neglected under the usual assumption $\Omega_g \ll \Omega_r$. Therefore we also optimize the parameters based the simplified three-level EIT consisting of the three levels $\{\vert g\rangle, \vert r\rangle, \vert e\rangle \}$ with $T\nu=1200$, and then substitute the optimal set of parameters back into the four-level EIT for real time evolution, which is shown in black solid line in Fig.~\ref{fig:fig4}(b). The optimal parameters found at different $T$s are also shown in Table.~\ref{tab:fourleveloptimal}, together with the cooling rates (denoted as $W$) from exponential fittings of the time evolution. The last row of Table.~\ref{tab:fourleveloptimal} shows the optimal parameters for the three-level EIT in which the energy level $\vert t\rangle$ is neglected, corresponding to the black line in Fig.~\ref{fig:fig4}(b). Different from the simulations done in Fig.~\ref{fig:fig3}, here $\Delta_g$ and $\Delta_r$ are treated independently but their optimal values are still equal to each other. The last column of Table.~\ref{tab:fourleveloptimal} shows $\nave(t)$ when $\nu t=1200$. We can see that fast cooling could still be achieved in the realistic four-level EIT scheme. In particular, we can reach a cooling rate of $W=0.015$, which means that ground state cooling can be achieved in several tens of periods of the trap frequency. Interestingly, in comparison with the standard three-level EIT, the optimal $\Delta_g$ and $\Delta_r$ in the four-level case are no longer equal, that is, $\Delta_g - \Delta_r \approx 0.1\nu$ at $T\nu=300,700, 1200$. Additionally, we can see that by direct optimizing the four-level scheme, one  obtain a parameter setting which is almost as optimal as that based on the simplified three-level scheme, which justifies the approximation to neglect the level $\vert t\rangle$. 

% We can see from the table that with the optimal values from quantum control, we could achieve both fast cooling as well as a lower final average phonon occupation than that predicted theoretically.

% \gcc{However, it is slightly different from standard EIT cooling, because
% that the detunings $\Delta_{g}$ and $\Delta_{r}$ are not chosen
% to the same. The optimal cooling condition should be 
% \[
% \left(\Delta_{g}+\nu\right)\left(\Delta_{gr}+\nu\right)-\dfrac{\Omega_{r}^{2}}{4}=0
% \]
% Physically, it corresponds to the cooling laser in resonance with
% red sideband transition, $\left|g\right\rangle \left|n\right\rangle \longrightarrow\left|+\right\rangle \left|n-1\right\rangle $,
% where the narrow-linewidth dressed state $\left|+\right\rangle $
% is formed by states $\left|e\right\rangle $ and $\left|r\right\rangle $
% together with the coupling laser.}

\section{Conclusion}\label{sec:summary}

To summarize, we have proposed to enhance the classical cooling schemes using optimal quantum control. We review the standard sideband cooling and EIT cooling schemes and recompute the optimal cooling conditions numerically using optimal quantum control. The resulting set of optimal parameters does not satisfy the weak sideband coupling condition in general, and can result in a much faster cooling speed compared to that from the theoretical optimal cooling conditions derived based on the weak sideband coupling condition. We show that a very low average phonon occupation comparable to that predicted theoretically based on the weak sideband coupling condition can still be retained in our scheme. Moreover, we show with the example of a realistic four-level EIT scheme that our approach could easily be used for cooling of trapped ions with more complicated setups which often happens in real experiments, where an analytic derivation of the optimal cooling condition could be extremely difficult. Our optimal quantum control enhanced cooling scheme could also be directly applied to study other systems with a similar form of parametric Lindbladian, such as cooling of trapped single atom~\cite{Thompson2013coherence,Kaufman2012cooling,reiserer2013ground} and nano mechanical resonantor~\cite{chan2011laser}.
% Our optimal quantum control enhanced cooling scheme could hopefully be used as an accomplishing tool for ion cooling experiments in the future.

\begin{acknowledgments}
We thank Shi-Lei Su for fruitful discussion. C. G acknowledges support from National Natural Science Foundation of China under Grants No. 11805279, No. 61833010, No. 12074117 and No. 12061131011. P. C acknowledges support from National Natural Science Foundation of China under Grants No. 12074433. J.Z acknowledges support from National Natural Science Foundation of China under grant No. 12004430. W. W acknowledges support from National Basic Research Program of China under Grant No. 2016YFA0301903.
\end{acknowledgments}

\bibliographystyle{apsrev4-1}
% \bibliography{refs}

\begin{thebibliography}{67}%
\makeatletter
\providecommand \@ifxundefined [1]{%
 \@ifx{#1\undefined}
}%
\providecommand \@ifnum [1]{%
 \ifnum #1\expandafter \@firstoftwo
 \else \expandafter \@secondoftwo
 \fi
}%
\providecommand \@ifx [1]{%
 \ifx #1\expandafter \@firstoftwo
 \else \expandafter \@secondoftwo
 \fi
}%
\providecommand \natexlab [1]{#1}%
\providecommand \enquote  [1]{``#1''}%
\providecommand \bibnamefont  [1]{#1}%
\providecommand \bibfnamefont [1]{#1}%
\providecommand \citenamefont [1]{#1}%
\providecommand \href@noop [0]{\@secondoftwo}%
\providecommand \href [0]{\begingroup \@sanitize@url \@href}%
\providecommand \@href[1]{\@@startlink{#1}\@@href}%
\providecommand \@@href[1]{\endgroup#1\@@endlink}%
\providecommand \@sanitize@url [0]{\catcode `\\12\catcode `\$12\catcode
  `\&12\catcode `\#12\catcode `\^12\catcode `\_12\catcode `\%12\relax}%
\providecommand \@@startlink[1]{}%
\providecommand \@@endlink[0]{}%
\providecommand \url  [0]{\begingroup\@sanitize@url \@url }%
\providecommand \@url [1]{\endgroup\@href {#1}{\urlprefix }}%
\providecommand \urlprefix  [0]{URL }%
\providecommand \Eprint [0]{\href }%
\providecommand \doibase [0]{http://dx.doi.org/}%
\providecommand \selectlanguage [0]{\@gobble}%
\providecommand \bibinfo  [0]{\@secondoftwo}%
\providecommand \bibfield  [0]{\@secondoftwo}%
\providecommand \translation [1]{[#1]}%
\providecommand \BibitemOpen [0]{}%
\providecommand \bibitemStop [0]{}%
\providecommand \bibitemNoStop [0]{.\EOS\space}%
\providecommand \EOS [0]{\spacefactor3000\relax}%
\providecommand \BibitemShut  [1]{\csname bibitem#1\endcsname}%
\let\auto@bib@innerbib\@empty
%</preamble>
\bibitem [{\citenamefont {Wineland}\ \emph {et~al.}(1998)\citenamefont
  {Wineland}, \citenamefont {Monroe}, \citenamefont {Itano}, \citenamefont
  {Leibfried}, \citenamefont {King},\ and\ \citenamefont
  {Meekhof}}]{WinelandMeekhof1998}%
  \BibitemOpen
  \bibfield  {author} {\bibinfo {author} {\bibfnamefont {D.~J.}\ \bibnamefont
  {Wineland}}, \bibinfo {author} {\bibfnamefont {C.}~\bibnamefont {Monroe}},
  \bibinfo {author} {\bibfnamefont {W.~M.}\ \bibnamefont {Itano}}, \bibinfo
  {author} {\bibfnamefont {D.}~\bibnamefont {Leibfried}}, \bibinfo {author}
  {\bibfnamefont {B.~E.}\ \bibnamefont {King}}, \ and\ \bibinfo {author}
  {\bibfnamefont {D.~M.}\ \bibnamefont {Meekhof}},\ }\href@noop {} {\bibfield
  {journal} {\bibinfo  {journal} {Journal of Research of the National Institute
  of Standards and Technology}\ }\textbf {\bibinfo {volume} {103}},\ \bibinfo
  {pages} {259} (\bibinfo {year} {1998})}\BibitemShut {NoStop}%
\bibitem [{\citenamefont {Porras}\ and\ \citenamefont
  {Cirac}(2004{\natexlab{a}})}]{PorrasCirac2004b}%
  \BibitemOpen
  \bibfield  {author} {\bibinfo {author} {\bibfnamefont {D.}~\bibnamefont
  {Porras}}\ and\ \bibinfo {author} {\bibfnamefont {J.~I.}\ \bibnamefont
  {Cirac}},\ }\href@noop {} {\bibfield  {journal} {\bibinfo  {journal}
  {Physical Review Letters}\ }\textbf {\bibinfo {volume} {92}},\ \bibinfo
  {pages} {207901} (\bibinfo {year} {2004}{\natexlab{a}})}\BibitemShut
  {NoStop}%
\bibitem [{\citenamefont {Porras}\ and\ \citenamefont
  {Cirac}(2004{\natexlab{b}})}]{PorrasCirac2004a}%
  \BibitemOpen
  \bibfield  {author} {\bibinfo {author} {\bibfnamefont {D.}~\bibnamefont
  {Porras}}\ and\ \bibinfo {author} {\bibfnamefont {J.~I.}\ \bibnamefont
  {Cirac}},\ }\href@noop {} {\bibfield  {journal} {\bibinfo  {journal}
  {Physical Review Letters}\ }\textbf {\bibinfo {volume} {93}},\ \bibinfo
  {pages} {263602} (\bibinfo {year} {2004}{\natexlab{b}})}\BibitemShut
  {NoStop}%
\bibitem [{\citenamefont {Leibfried}\ \emph {et~al.}(2003)\citenamefont
  {Leibfried}, \citenamefont {Blatt}, \citenamefont {Monroe},\ and\
  \citenamefont {Wineland}}]{LeibfriedWineland2003}%
  \BibitemOpen
  \bibfield  {author} {\bibinfo {author} {\bibfnamefont {D.}~\bibnamefont
  {Leibfried}}, \bibinfo {author} {\bibfnamefont {R.}~\bibnamefont {Blatt}},
  \bibinfo {author} {\bibfnamefont {C.}~\bibnamefont {Monroe}}, \ and\ \bibinfo
  {author} {\bibfnamefont {D.}~\bibnamefont {Wineland}},\ }\href@noop {}
  {\bibfield  {journal} {\bibinfo  {journal} {Reviews of Modern Physics}\
  }\textbf {\bibinfo {volume} {75}},\ \bibinfo {pages} {281} (\bibinfo {year}
  {2003})}\BibitemShut {NoStop}%
\bibitem [{\citenamefont {Berm{\'u}dez}\ \emph {et~al.}(2013)\citenamefont
  {Berm{\'u}dez}, \citenamefont {Bruderer},\ and\ \citenamefont
  {Plenio}}]{BermudezPlenio2013}%
  \BibitemOpen
  \bibfield  {author} {\bibinfo {author} {\bibfnamefont {A.}~\bibnamefont
  {Berm{\'u}dez}}, \bibinfo {author} {\bibfnamefont {M.}~\bibnamefont
  {Bruderer}}, \ and\ \bibinfo {author} {\bibfnamefont {M.~B.}\ \bibnamefont
  {Plenio}},\ }\href@noop {} {\bibfield  {journal} {\bibinfo  {journal}
  {Physical Review Letters}\ }\textbf {\bibinfo {volume} {111}},\ \bibinfo
  {pages} {040601} (\bibinfo {year} {2013})}\BibitemShut {NoStop}%
\bibitem [{\citenamefont {Ruiz}\ \emph {et~al.}(2014)\citenamefont {Ruiz},
  \citenamefont {Alonso}, \citenamefont {Plenio},\ and\ \citenamefont {del
  Campo}}]{RuizCampo2014}%
  \BibitemOpen
  \bibfield  {author} {\bibinfo {author} {\bibfnamefont {A.}~\bibnamefont
  {Ruiz}}, \bibinfo {author} {\bibfnamefont {D.}~\bibnamefont {Alonso}},
  \bibinfo {author} {\bibfnamefont {M.~B.}\ \bibnamefont {Plenio}}, \ and\
  \bibinfo {author} {\bibfnamefont {A.}~\bibnamefont {del Campo}},\ }\href@noop
  {} {\bibfield  {journal} {\bibinfo  {journal} {Physical Review B}\ }\textbf
  {\bibinfo {volume} {89}},\ \bibinfo {pages} {214305} (\bibinfo {year}
  {2014})}\BibitemShut {NoStop}%
\bibitem [{\citenamefont {Ramm}\ \emph {et~al.}(2014)\citenamefont {Ramm},
  \citenamefont {Pruttivarasin},\ and\ \citenamefont
  {H{\"a}ffner}}]{RammHartmut2014}%
  \BibitemOpen
  \bibfield  {author} {\bibinfo {author} {\bibfnamefont {M.}~\bibnamefont
  {Ramm}}, \bibinfo {author} {\bibfnamefont {T.}~\bibnamefont {Pruttivarasin}},
  \ and\ \bibinfo {author} {\bibfnamefont {H.}~\bibnamefont {H{\"a}ffner}},\
  }\href@noop {} {\bibfield  {journal} {\bibinfo  {journal} {New Journal of
  Physics}\ }\textbf {\bibinfo {volume} {16}},\ \bibinfo {pages} {063062}
  (\bibinfo {year} {2014})}\BibitemShut {NoStop}%
\bibitem [{\citenamefont {Guo}\ \emph {et~al.}(2015)\citenamefont {Guo},
  \citenamefont {Mukherjee},\ and\ \citenamefont {Poletti}}]{GuoPoletti2015}%
  \BibitemOpen
  \bibfield  {author} {\bibinfo {author} {\bibfnamefont {C.}~\bibnamefont
  {Guo}}, \bibinfo {author} {\bibfnamefont {M.}~\bibnamefont {Mukherjee}}, \
  and\ \bibinfo {author} {\bibfnamefont {D.}~\bibnamefont {Poletti}},\
  }\href@noop {} {\bibfield  {journal} {\bibinfo  {journal} {Physical Review
  A}\ }\textbf {\bibinfo {volume} {92}},\ \bibinfo {pages} {023637} (\bibinfo
  {year} {2015})}\BibitemShut {NoStop}%
\bibitem [{\citenamefont {Guo}\ and\ \citenamefont
  {Poletti}(2016)}]{GuoPoletti2016}%
  \BibitemOpen
  \bibfield  {author} {\bibinfo {author} {\bibfnamefont {C.}~\bibnamefont
  {Guo}}\ and\ \bibinfo {author} {\bibfnamefont {D.}~\bibnamefont {Poletti}},\
  }\href@noop {} {\bibfield  {journal} {\bibinfo  {journal} {Physical Review
  A}\ }\textbf {\bibinfo {volume} {94}},\ \bibinfo {pages} {033610} (\bibinfo
  {year} {2016})}\BibitemShut {NoStop}%
\bibitem [{\citenamefont {Guo}\ and\ \citenamefont
  {Poletti}(2017{\natexlab{a}})}]{GuoPoletti2017}%
  \BibitemOpen
  \bibfield  {author} {\bibinfo {author} {\bibfnamefont {C.}~\bibnamefont
  {Guo}}\ and\ \bibinfo {author} {\bibfnamefont {D.}~\bibnamefont {Poletti}},\
  }\href@noop {} {\bibfield  {journal} {\bibinfo  {journal} {Physical Review
  A}\ }\textbf {\bibinfo {volume} {95}},\ \bibinfo {pages} {052107} (\bibinfo
  {year} {2017}{\natexlab{a}})}\BibitemShut {NoStop}%
\bibitem [{\citenamefont {Guo}\ and\ \citenamefont
  {Poletti}(2017{\natexlab{b}})}]{GuoPoletti2017b}%
  \BibitemOpen
  \bibfield  {author} {\bibinfo {author} {\bibfnamefont {C.}~\bibnamefont
  {Guo}}\ and\ \bibinfo {author} {\bibfnamefont {D.}~\bibnamefont {Poletti}},\
  }\href@noop {} {\bibfield  {journal} {\bibinfo  {journal} {Physical Review
  B}\ }\textbf {\bibinfo {volume} {96}},\ \bibinfo {pages} {165409} (\bibinfo
  {year} {2017}{\natexlab{b}})}\BibitemShut {NoStop}%
\bibitem [{\citenamefont {Guo}\ and\ \citenamefont
  {Poletti}(2018)}]{GuoPoletti2018}%
  \BibitemOpen
  \bibfield  {author} {\bibinfo {author} {\bibfnamefont {C.}~\bibnamefont
  {Guo}}\ and\ \bibinfo {author} {\bibfnamefont {D.}~\bibnamefont {Poletti}},\
  }\href@noop {} {\bibfield  {journal} {\bibinfo  {journal} {Physical Review
  A}\ }\textbf {\bibinfo {volume} {98}},\ \bibinfo {pages} {052126} (\bibinfo
  {year} {2018})}\BibitemShut {NoStop}%
\bibitem [{\citenamefont {Xu}\ \emph {et~al.}(2019)\citenamefont {Xu},
  \citenamefont {Thingna}, \citenamefont {Guo},\ and\ \citenamefont
  {Poletti}}]{XuPoletti2019}%
  \BibitemOpen
  \bibfield  {author} {\bibinfo {author} {\bibfnamefont {X.}~\bibnamefont
  {Xu}}, \bibinfo {author} {\bibfnamefont {J.}~\bibnamefont {Thingna}},
  \bibinfo {author} {\bibfnamefont {C.}~\bibnamefont {Guo}}, \ and\ \bibinfo
  {author} {\bibfnamefont {D.}~\bibnamefont {Poletti}},\ }\href@noop {}
  {\bibfield  {journal} {\bibinfo  {journal} {Physical Review A}\ }\textbf
  {\bibinfo {volume} {99}},\ \bibinfo {pages} {012106} (\bibinfo {year}
  {2019})}\BibitemShut {NoStop}%
\bibitem [{\citenamefont {Pan}\ \emph {et~al.}(2020)\citenamefont {Pan},
  \citenamefont {Zhang}, \citenamefont {Cohen}, \citenamefont {Wu},
  \citenamefont {Chen},\ and\ \citenamefont {Davidson}}]{PanDavidson2020}%
  \BibitemOpen
  \bibfield  {author} {\bibinfo {author} {\bibfnamefont {Y.}~\bibnamefont
  {Pan}}, \bibinfo {author} {\bibfnamefont {J.}~\bibnamefont {Zhang}}, \bibinfo
  {author} {\bibfnamefont {E.}~\bibnamefont {Cohen}}, \bibinfo {author}
  {\bibfnamefont {C.-w.}\ \bibnamefont {Wu}}, \bibinfo {author} {\bibfnamefont
  {P.-X.}\ \bibnamefont {Chen}}, \ and\ \bibinfo {author} {\bibfnamefont
  {N.}~\bibnamefont {Davidson}},\ }\href@noop {} {\bibfield  {journal}
  {\bibinfo  {journal} {Nature Physics}\ }\textbf {\bibinfo {volume} {16}},\
  \bibinfo {pages} {1206} (\bibinfo {year} {2020})}\BibitemShut {NoStop}%
\bibitem [{\citenamefont {Wu}\ \emph {et~al.}(2019)\citenamefont {Wu},
  \citenamefont {Zhang}, \citenamefont {Xie}, \citenamefont {Ou}, \citenamefont
  {Chen}, \citenamefont {Wu},\ and\ \citenamefont {Chen}}]{WuChen2019}%
  \BibitemOpen
  \bibfield  {author} {\bibinfo {author} {\bibfnamefont {C.-W.}\ \bibnamefont
  {Wu}}, \bibinfo {author} {\bibfnamefont {J.}~\bibnamefont {Zhang}}, \bibinfo
  {author} {\bibfnamefont {Y.}~\bibnamefont {Xie}}, \bibinfo {author}
  {\bibfnamefont {B.-Q.}\ \bibnamefont {Ou}}, \bibinfo {author} {\bibfnamefont
  {T.}~\bibnamefont {Chen}}, \bibinfo {author} {\bibfnamefont {W.}~\bibnamefont
  {Wu}}, \ and\ \bibinfo {author} {\bibfnamefont {P.-X.}\ \bibnamefont
  {Chen}},\ }\href@noop {} {\bibfield  {journal} {\bibinfo  {journal} {Physical
  Review A}\ }\textbf {\bibinfo {volume} {100}},\ \bibinfo {pages} {062111}
  (\bibinfo {year} {2019})}\BibitemShut {NoStop}%
\bibitem [{\citenamefont {Cirac}\ and\ \citenamefont
  {Zoller}(1995)}]{CiracZoller1995}%
  \BibitemOpen
  \bibfield  {author} {\bibinfo {author} {\bibfnamefont {J.~I.}\ \bibnamefont
  {Cirac}}\ and\ \bibinfo {author} {\bibfnamefont {P.}~\bibnamefont {Zoller}},\
  }\href@noop {} {\bibfield  {journal} {\bibinfo  {journal} {Physical Review
  Letters}\ }\textbf {\bibinfo {volume} {74}},\ \bibinfo {pages} {4091}
  (\bibinfo {year} {1995})}\BibitemShut {NoStop}%
\bibitem [{\citenamefont {Lanyon}\ \emph {et~al.}(2011)\citenamefont {Lanyon},
  \citenamefont {Hempel}, \citenamefont {Nigg}, \citenamefont {M{\"u}ller},
  \citenamefont {Gerritsma}, \citenamefont {Z{\"a}hringer}, \citenamefont
  {Schindler}, \citenamefont {Barreiro}, \citenamefont {Rambach}, \citenamefont
  {Kirchmair} \emph {et~al.}}]{LanyonRoos2011}%
  \BibitemOpen
  \bibfield  {author} {\bibinfo {author} {\bibfnamefont {B.~P.}\ \bibnamefont
  {Lanyon}}, \bibinfo {author} {\bibfnamefont {C.}~\bibnamefont {Hempel}},
  \bibinfo {author} {\bibfnamefont {D.}~\bibnamefont {Nigg}}, \bibinfo {author}
  {\bibfnamefont {M.}~\bibnamefont {M{\"u}ller}}, \bibinfo {author}
  {\bibfnamefont {R.}~\bibnamefont {Gerritsma}}, \bibinfo {author}
  {\bibfnamefont {F.}~\bibnamefont {Z{\"a}hringer}}, \bibinfo {author}
  {\bibfnamefont {P.}~\bibnamefont {Schindler}}, \bibinfo {author}
  {\bibfnamefont {J.~T.}\ \bibnamefont {Barreiro}}, \bibinfo {author}
  {\bibfnamefont {M.}~\bibnamefont {Rambach}}, \bibinfo {author} {\bibfnamefont
  {G.}~\bibnamefont {Kirchmair}},  \emph {et~al.},\ }\href@noop {} {\bibfield
  {journal} {\bibinfo  {journal} {Science}\ }\textbf {\bibinfo {volume}
  {334}},\ \bibinfo {pages} {57} (\bibinfo {year} {2011})}\BibitemShut
  {NoStop}%
\bibitem [{\citenamefont {Kielpinski}\ \emph {et~al.}(2002)\citenamefont
  {Kielpinski}, \citenamefont {Monroe},\ and\ \citenamefont
  {Wineland}}]{KielpinskiWineland2002}%
  \BibitemOpen
  \bibfield  {author} {\bibinfo {author} {\bibfnamefont {D.}~\bibnamefont
  {Kielpinski}}, \bibinfo {author} {\bibfnamefont {C.}~\bibnamefont {Monroe}},
  \ and\ \bibinfo {author} {\bibfnamefont {D.~J.}\ \bibnamefont {Wineland}},\
  }\href@noop {} {\bibfield  {journal} {\bibinfo  {journal} {Nature}\ }\textbf
  {\bibinfo {volume} {417}},\ \bibinfo {pages} {709} (\bibinfo {year}
  {2002})}\BibitemShut {NoStop}%
\bibitem [{\citenamefont {Diedrich}\ \emph {et~al.}(1989)\citenamefont
  {Diedrich}, \citenamefont {Bergquist}, \citenamefont {Itano},\ and\
  \citenamefont {Wineland}}]{DiedrichWineland1989}%
  \BibitemOpen
  \bibfield  {author} {\bibinfo {author} {\bibfnamefont {F.}~\bibnamefont
  {Diedrich}}, \bibinfo {author} {\bibfnamefont {J.}~\bibnamefont {Bergquist}},
  \bibinfo {author} {\bibfnamefont {W.~M.}\ \bibnamefont {Itano}}, \ and\
  \bibinfo {author} {\bibfnamefont {D.}~\bibnamefont {Wineland}},\ }\href@noop
  {} {\bibfield  {journal} {\bibinfo  {journal} {Physical Review Letters}\
  }\textbf {\bibinfo {volume} {62}},\ \bibinfo {pages} {403} (\bibinfo {year}
  {1989})}\BibitemShut {NoStop}%
\bibitem [{\citenamefont {Cirac}\ \emph {et~al.}(1992)\citenamefont {Cirac},
  \citenamefont {Blatt}, \citenamefont {Zoller},\ and\ \citenamefont
  {Phillips}}]{CiracZoller1992}%
  \BibitemOpen
  \bibfield  {author} {\bibinfo {author} {\bibfnamefont {J.~I.}\ \bibnamefont
  {Cirac}}, \bibinfo {author} {\bibfnamefont {R.}~\bibnamefont {Blatt}},
  \bibinfo {author} {\bibfnamefont {P.}~\bibnamefont {Zoller}}, \ and\ \bibinfo
  {author} {\bibfnamefont {W.~D.}\ \bibnamefont {Phillips}},\ }\href@noop {}
  {\bibfield  {journal} {\bibinfo  {journal} {Physical Review A}\ }\textbf
  {\bibinfo {volume} {46}},\ \bibinfo {pages} {2668} (\bibinfo {year}
  {1992})}\BibitemShut {NoStop}%
\bibitem [{\citenamefont {Monroe}\ \emph {et~al.}(1995)\citenamefont {Monroe},
  \citenamefont {Meekhof}, \citenamefont {King}, \citenamefont {Jefferts},
  \citenamefont {Itano}, \citenamefont {Wineland},\ and\ \citenamefont
  {Gould}}]{MonroeWineland1995}%
  \BibitemOpen
  \bibfield  {author} {\bibinfo {author} {\bibfnamefont {C.}~\bibnamefont
  {Monroe}}, \bibinfo {author} {\bibfnamefont {D.}~\bibnamefont {Meekhof}},
  \bibinfo {author} {\bibfnamefont {B.}~\bibnamefont {King}}, \bibinfo {author}
  {\bibfnamefont {S.~R.}\ \bibnamefont {Jefferts}}, \bibinfo {author}
  {\bibfnamefont {W.~M.}\ \bibnamefont {Itano}}, \bibinfo {author}
  {\bibfnamefont {D.~J.}\ \bibnamefont {Wineland}}, \ and\ \bibinfo {author}
  {\bibfnamefont {P.}~\bibnamefont {Gould}},\ }\href@noop {} {\bibfield
  {journal} {\bibinfo  {journal} {Physical Review Letters}\ }\textbf {\bibinfo
  {volume} {75}},\ \bibinfo {pages} {4011} (\bibinfo {year}
  {1995})}\BibitemShut {NoStop}%
\bibitem [{\citenamefont {Roos}\ \emph {et~al.}(1999)\citenamefont {Roos},
  \citenamefont {Zeiger}, \citenamefont {Rohde}, \citenamefont {N{\"a}gerl},
  \citenamefont {Eschner}, \citenamefont {Leibfried}, \citenamefont
  {Schmidt-Kaler},\ and\ \citenamefont {Blatt}}]{RoosBlatt1999}%
  \BibitemOpen
  \bibfield  {author} {\bibinfo {author} {\bibfnamefont {C.}~\bibnamefont
  {Roos}}, \bibinfo {author} {\bibfnamefont {T.}~\bibnamefont {Zeiger}},
  \bibinfo {author} {\bibfnamefont {H.}~\bibnamefont {Rohde}}, \bibinfo
  {author} {\bibfnamefont {H.}~\bibnamefont {N{\"a}gerl}}, \bibinfo {author}
  {\bibfnamefont {J.}~\bibnamefont {Eschner}}, \bibinfo {author} {\bibfnamefont
  {D.}~\bibnamefont {Leibfried}}, \bibinfo {author} {\bibfnamefont
  {F.}~\bibnamefont {Schmidt-Kaler}}, \ and\ \bibinfo {author} {\bibfnamefont
  {R.}~\bibnamefont {Blatt}},\ }\href@noop {} {\bibfield  {journal} {\bibinfo
  {journal} {Physical Review Letters}\ }\textbf {\bibinfo {volume} {83}},\
  \bibinfo {pages} {4713} (\bibinfo {year} {1999})}\BibitemShut {NoStop}%
\bibitem [{\citenamefont {Morigi}\ \emph {et~al.}(2000)\citenamefont {Morigi},
  \citenamefont {Eschner},\ and\ \citenamefont {Keitel}}]{MorigiKeitel2000}%
  \BibitemOpen
  \bibfield  {author} {\bibinfo {author} {\bibfnamefont {G.}~\bibnamefont
  {Morigi}}, \bibinfo {author} {\bibfnamefont {J.}~\bibnamefont {Eschner}}, \
  and\ \bibinfo {author} {\bibfnamefont {C.~H.}\ \bibnamefont {Keitel}},\
  }\href@noop {} {\bibfield  {journal} {\bibinfo  {journal} {Physical Review
  Letters}\ }\textbf {\bibinfo {volume} {85}},\ \bibinfo {pages} {4458}
  (\bibinfo {year} {2000})}\BibitemShut {NoStop}%
\bibitem [{\citenamefont {Roos}\ \emph {et~al.}(2000)\citenamefont {Roos},
  \citenamefont {Leibfried}, \citenamefont {Mundt}, \citenamefont
  {Schmidt-Kaler}, \citenamefont {Eschner},\ and\ \citenamefont
  {Blatt}}]{RoosBlatt2000}%
  \BibitemOpen
  \bibfield  {author} {\bibinfo {author} {\bibfnamefont {C.}~\bibnamefont
  {Roos}}, \bibinfo {author} {\bibfnamefont {D.}~\bibnamefont {Leibfried}},
  \bibinfo {author} {\bibfnamefont {A.}~\bibnamefont {Mundt}}, \bibinfo
  {author} {\bibfnamefont {F.}~\bibnamefont {Schmidt-Kaler}}, \bibinfo {author}
  {\bibfnamefont {J.}~\bibnamefont {Eschner}}, \ and\ \bibinfo {author}
  {\bibfnamefont {R.}~\bibnamefont {Blatt}},\ }\href@noop {} {\bibfield
  {journal} {\bibinfo  {journal} {Physical Review Letters}\ }\textbf {\bibinfo
  {volume} {85}},\ \bibinfo {pages} {5547} (\bibinfo {year}
  {2000})}\BibitemShut {NoStop}%
\bibitem [{\citenamefont {Evers}\ and\ \citenamefont
  {Keitel}(2004)}]{EversKeitel2004}%
  \BibitemOpen
  \bibfield  {author} {\bibinfo {author} {\bibfnamefont {J.}~\bibnamefont
  {Evers}}\ and\ \bibinfo {author} {\bibfnamefont {C.~H.}\ \bibnamefont
  {Keitel}},\ }\href@noop {} {\bibfield  {journal} {\bibinfo  {journal} {EPL
  (Europhysics Letters)}\ }\textbf {\bibinfo {volume} {68}},\ \bibinfo {pages}
  {370} (\bibinfo {year} {2004})}\BibitemShut {NoStop}%
\bibitem [{\citenamefont {Retzker}\ and\ \citenamefont
  {Plenio}(2007)}]{RetzkerPlenio2007}%
  \BibitemOpen
  \bibfield  {author} {\bibinfo {author} {\bibfnamefont {A.}~\bibnamefont
  {Retzker}}\ and\ \bibinfo {author} {\bibfnamefont {M.}~\bibnamefont
  {Plenio}},\ }\href@noop {} {\bibfield  {journal} {\bibinfo  {journal} {New
  Journal of Physics}\ }\textbf {\bibinfo {volume} {9}},\ \bibinfo {pages}
  {279} (\bibinfo {year} {2007})}\BibitemShut {NoStop}%
\bibitem [{\citenamefont {Cerrillo}\ \emph {et~al.}(2010)\citenamefont
  {Cerrillo}, \citenamefont {Retzker},\ and\ \citenamefont
  {Plenio}}]{CerrilloPlenio2010}%
  \BibitemOpen
  \bibfield  {author} {\bibinfo {author} {\bibfnamefont {J.}~\bibnamefont
  {Cerrillo}}, \bibinfo {author} {\bibfnamefont {A.}~\bibnamefont {Retzker}}, \
  and\ \bibinfo {author} {\bibfnamefont {M.~B.}\ \bibnamefont {Plenio}},\
  }\href@noop {} {\bibfield  {journal} {\bibinfo  {journal} {Physical Review
  Letters}\ }\textbf {\bibinfo {volume} {104}},\ \bibinfo {pages} {043003}
  (\bibinfo {year} {2010})}\BibitemShut {NoStop}%
\bibitem [{\citenamefont {Albrecht}\ \emph {et~al.}(2011)\citenamefont
  {Albrecht}, \citenamefont {Retzker}, \citenamefont {Wunderlich},\ and\
  \citenamefont {Plenio}}]{AlbrechtPlenio2011}%
  \BibitemOpen
  \bibfield  {author} {\bibinfo {author} {\bibfnamefont {A.}~\bibnamefont
  {Albrecht}}, \bibinfo {author} {\bibfnamefont {A.}~\bibnamefont {Retzker}},
  \bibinfo {author} {\bibfnamefont {C.}~\bibnamefont {Wunderlich}}, \ and\
  \bibinfo {author} {\bibfnamefont {M.~B.}\ \bibnamefont {Plenio}},\
  }\href@noop {} {\bibfield  {journal} {\bibinfo  {journal} {New Journal of
  Physics}\ }\textbf {\bibinfo {volume} {13}},\ \bibinfo {pages} {033009}
  (\bibinfo {year} {2011})}\BibitemShut {NoStop}%
\bibitem [{\citenamefont {Zhang}\ \emph {et~al.}(2012)\citenamefont {Zhang},
  \citenamefont {Wu},\ and\ \citenamefont {Chen}}]{ZhangChen2012}%
  \BibitemOpen
  \bibfield  {author} {\bibinfo {author} {\bibfnamefont {S.}~\bibnamefont
  {Zhang}}, \bibinfo {author} {\bibfnamefont {C.-W.}\ \bibnamefont {Wu}}, \
  and\ \bibinfo {author} {\bibfnamefont {P.-X.}\ \bibnamefont {Chen}},\
  }\href@noop {} {\bibfield  {journal} {\bibinfo  {journal} {Physical Review
  A}\ }\textbf {\bibinfo {volume} {85}},\ \bibinfo {pages} {053420} (\bibinfo
  {year} {2012})}\BibitemShut {NoStop}%
\bibitem [{\citenamefont {Yi}\ \emph {et~al.}(2013)\citenamefont {Yi},
  \citenamefont {Li},\ and\ \citenamefont {Yang}}]{YiYang2013}%
  \BibitemOpen
  \bibfield  {author} {\bibinfo {author} {\bibfnamefont {Z.}~\bibnamefont
  {Yi}}, \bibinfo {author} {\bibfnamefont {G.-x.}\ \bibnamefont {Li}}, \ and\
  \bibinfo {author} {\bibfnamefont {Y.-p.}\ \bibnamefont {Yang}},\ }\href@noop
  {} {\bibfield  {journal} {\bibinfo  {journal} {Physical Review A}\ }\textbf
  {\bibinfo {volume} {87}},\ \bibinfo {pages} {053408} (\bibinfo {year}
  {2013})}\BibitemShut {NoStop}%
\bibitem [{\citenamefont {Zhang}\ \emph {et~al.}(2014)\citenamefont {Zhang},
  \citenamefont {Duan}, \citenamefont {Guo}, \citenamefont {Wu}, \citenamefont
  {Wu},\ and\ \citenamefont {Chen}}]{ZhangChen2014}%
  \BibitemOpen
  \bibfield  {author} {\bibinfo {author} {\bibfnamefont {S.}~\bibnamefont
  {Zhang}}, \bibinfo {author} {\bibfnamefont {Q.-H.}\ \bibnamefont {Duan}},
  \bibinfo {author} {\bibfnamefont {C.}~\bibnamefont {Guo}}, \bibinfo {author}
  {\bibfnamefont {C.-W.}\ \bibnamefont {Wu}}, \bibinfo {author} {\bibfnamefont
  {W.}~\bibnamefont {Wu}}, \ and\ \bibinfo {author} {\bibfnamefont {P.-X.}\
  \bibnamefont {Chen}},\ }\href@noop {} {\bibfield  {journal} {\bibinfo
  {journal} {Physical Review A}\ }\textbf {\bibinfo {volume} {89}},\ \bibinfo
  {pages} {013402} (\bibinfo {year} {2014})}\BibitemShut {NoStop}%
\bibitem [{\citenamefont {Lu}\ \emph {et~al.}(2015)\citenamefont {Lu},
  \citenamefont {Zhang}, \citenamefont {Cui}, \citenamefont {Cao},
  \citenamefont {Zhang}, \citenamefont {Huang}, \citenamefont {Li},\ and\
  \citenamefont {Guo}}]{LuGuo2015}%
  \BibitemOpen
  \bibfield  {author} {\bibinfo {author} {\bibfnamefont {Y.}~\bibnamefont
  {Lu}}, \bibinfo {author} {\bibfnamefont {J.-Q.}\ \bibnamefont {Zhang}},
  \bibinfo {author} {\bibfnamefont {J.-M.}\ \bibnamefont {Cui}}, \bibinfo
  {author} {\bibfnamefont {D.-Y.}\ \bibnamefont {Cao}}, \bibinfo {author}
  {\bibfnamefont {S.}~\bibnamefont {Zhang}}, \bibinfo {author} {\bibfnamefont
  {Y.-F.}\ \bibnamefont {Huang}}, \bibinfo {author} {\bibfnamefont {C.-F.}\
  \bibnamefont {Li}}, \ and\ \bibinfo {author} {\bibfnamefont {G.-C.}\
  \bibnamefont {Guo}},\ }\href@noop {} {\bibfield  {journal} {\bibinfo
  {journal} {Physical Review A}\ }\textbf {\bibinfo {volume} {92}},\ \bibinfo
  {pages} {023420} (\bibinfo {year} {2015})}\BibitemShut {NoStop}%
\bibitem [{\citenamefont {Yi}\ and\ \citenamefont {Gu}(2017)}]{YiGu2017}%
  \BibitemOpen
  \bibfield  {author} {\bibinfo {author} {\bibfnamefont {Z.}~\bibnamefont
  {Yi}}\ and\ \bibinfo {author} {\bibfnamefont {W.-j.}\ \bibnamefont {Gu}},\
  }\href@noop {} {\bibfield  {journal} {\bibinfo  {journal} {Optics express}\
  }\textbf {\bibinfo {volume} {25}},\ \bibinfo {pages} {1314} (\bibinfo {year}
  {2017})}\BibitemShut {NoStop}%
\bibitem [{\citenamefont {Cerrillo}\ \emph {et~al.}(2018)\citenamefont
  {Cerrillo}, \citenamefont {Retzker},\ and\ \citenamefont
  {Plenio}}]{CerrilloPlenio2018}%
  \BibitemOpen
  \bibfield  {author} {\bibinfo {author} {\bibfnamefont {J.}~\bibnamefont
  {Cerrillo}}, \bibinfo {author} {\bibfnamefont {A.}~\bibnamefont {Retzker}}, \
  and\ \bibinfo {author} {\bibfnamefont {M.~B.}\ \bibnamefont {Plenio}},\
  }\href@noop {} {\bibfield  {journal} {\bibinfo  {journal} {Physical Review
  A}\ }\textbf {\bibinfo {volume} {98}},\ \bibinfo {pages} {013423} (\bibinfo
  {year} {2018})}\BibitemShut {NoStop}%
\bibitem [{\citenamefont {Zhang}\ \emph
  {et~al.}(2021{\natexlab{a}})\citenamefont {Zhang}, \citenamefont {Tian},
  \citenamefont {Wu}, \citenamefont {Zhang}, \citenamefont {Wang},
  \citenamefont {Wu}, \citenamefont {Bao},\ and\ \citenamefont
  {Guo}}]{ZhangGuo2021b}%
  \BibitemOpen
  \bibfield  {author} {\bibinfo {author} {\bibfnamefont {S.}~\bibnamefont
  {Zhang}}, \bibinfo {author} {\bibfnamefont {T.-C.}\ \bibnamefont {Tian}},
  \bibinfo {author} {\bibfnamefont {Z.-Y.}\ \bibnamefont {Wu}}, \bibinfo
  {author} {\bibfnamefont {Z.-S.}\ \bibnamefont {Zhang}}, \bibinfo {author}
  {\bibfnamefont {X.-H.}\ \bibnamefont {Wang}}, \bibinfo {author}
  {\bibfnamefont {W.}~\bibnamefont {Wu}}, \bibinfo {author} {\bibfnamefont
  {W.-S.}\ \bibnamefont {Bao}}, \ and\ \bibinfo {author} {\bibfnamefont
  {C.}~\bibnamefont {Guo}},\ }\href@noop {} {\bibfield  {journal} {\bibinfo
  {journal} {Physical Review A}\ }\textbf {\bibinfo {volume} {104}},\ \bibinfo
  {pages} {013117} (\bibinfo {year} {2021}{\natexlab{a}})}\BibitemShut
  {NoStop}%
\bibitem [{\citenamefont {Zhang}\ \emph
  {et~al.}(2021{\natexlab{b}})\citenamefont {Zhang}, \citenamefont {Zhang},
  \citenamefont {Wu}, \citenamefont {Bao},\ and\ \citenamefont
  {Guo}}]{ZhangGuo2021}%
  \BibitemOpen
  \bibfield  {author} {\bibinfo {author} {\bibfnamefont {S.}~\bibnamefont
  {Zhang}}, \bibinfo {author} {\bibfnamefont {J.-Q.}\ \bibnamefont {Zhang}},
  \bibinfo {author} {\bibfnamefont {W.}~\bibnamefont {Wu}}, \bibinfo {author}
  {\bibfnamefont {W.-S.}\ \bibnamefont {Bao}}, \ and\ \bibinfo {author}
  {\bibfnamefont {C.}~\bibnamefont {Guo}},\ }\href@noop {} {\bibfield
  {journal} {\bibinfo  {journal} {New Journal of Physics}\ } (\bibinfo {year}
  {2021}{\natexlab{b}})}\BibitemShut {NoStop}%
\bibitem [{\citenamefont {Machnes}\ \emph {et~al.}(2010)\citenamefont
  {Machnes}, \citenamefont {Plenio}, \citenamefont {Reznik}, \citenamefont
  {Steane},\ and\ \citenamefont {Retzker}}]{MachnesRetzker2010}%
  \BibitemOpen
  \bibfield  {author} {\bibinfo {author} {\bibfnamefont {S.}~\bibnamefont
  {Machnes}}, \bibinfo {author} {\bibfnamefont {M.~B.}\ \bibnamefont {Plenio}},
  \bibinfo {author} {\bibfnamefont {B.}~\bibnamefont {Reznik}}, \bibinfo
  {author} {\bibfnamefont {A.}~\bibnamefont {Steane}}, \ and\ \bibinfo {author}
  {\bibfnamefont {A.}~\bibnamefont {Retzker}},\ }\href@noop {} {\bibfield
  {journal} {\bibinfo  {journal} {Physical Review Letters}\ }\textbf {\bibinfo
  {volume} {104}},\ \bibinfo {pages} {183001} (\bibinfo {year}
  {2010})}\BibitemShut {NoStop}%
\bibitem [{\citenamefont {Wang}\ \emph {et~al.}(2011)\citenamefont {Wang},
  \citenamefont {Vinjanampathy}, \citenamefont {Strauch},\ and\ \citenamefont
  {Jacobs}}]{WangJacobs2011}%
  \BibitemOpen
  \bibfield  {author} {\bibinfo {author} {\bibfnamefont {X.}~\bibnamefont
  {Wang}}, \bibinfo {author} {\bibfnamefont {S.}~\bibnamefont {Vinjanampathy}},
  \bibinfo {author} {\bibfnamefont {F.~W.}\ \bibnamefont {Strauch}}, \ and\
  \bibinfo {author} {\bibfnamefont {K.}~\bibnamefont {Jacobs}},\ }\href@noop {}
  {\bibfield  {journal} {\bibinfo  {journal} {Physical review letters}\
  }\textbf {\bibinfo {volume} {107}},\ \bibinfo {pages} {177204} (\bibinfo
  {year} {2011})}\BibitemShut {NoStop}%
\bibitem [{\citenamefont {Machnes}\ \emph {et~al.}(2012)\citenamefont
  {Machnes}, \citenamefont {Cerrillo}, \citenamefont {Aspelmeyer},
  \citenamefont {Wieczorek}, \citenamefont {Plenio},\ and\ \citenamefont
  {Retzker}}]{MachnesRetzker2012}%
  \BibitemOpen
  \bibfield  {author} {\bibinfo {author} {\bibfnamefont {S.}~\bibnamefont
  {Machnes}}, \bibinfo {author} {\bibfnamefont {J.}~\bibnamefont {Cerrillo}},
  \bibinfo {author} {\bibfnamefont {M.}~\bibnamefont {Aspelmeyer}}, \bibinfo
  {author} {\bibfnamefont {W.}~\bibnamefont {Wieczorek}}, \bibinfo {author}
  {\bibfnamefont {M.~B.}\ \bibnamefont {Plenio}}, \ and\ \bibinfo {author}
  {\bibfnamefont {A.}~\bibnamefont {Retzker}},\ }\href@noop {} {\bibfield
  {journal} {\bibinfo  {journal} {Physical review letters}\ }\textbf {\bibinfo
  {volume} {108}},\ \bibinfo {pages} {153601} (\bibinfo {year}
  {2012})}\BibitemShut {NoStop}%
\bibitem [{\citenamefont {Liu}\ \emph {et~al.}(2013)\citenamefont {Liu},
  \citenamefont {Xiao}, \citenamefont {Luan},\ and\ \citenamefont
  {Wong}}]{LiuWong2013}%
  \BibitemOpen
  \bibfield  {author} {\bibinfo {author} {\bibfnamefont {Y.-C.}\ \bibnamefont
  {Liu}}, \bibinfo {author} {\bibfnamefont {Y.-F.}\ \bibnamefont {Xiao}},
  \bibinfo {author} {\bibfnamefont {X.}~\bibnamefont {Luan}}, \ and\ \bibinfo
  {author} {\bibfnamefont {C.~W.}\ \bibnamefont {Wong}},\ }\href@noop {}
  {\bibfield  {journal} {\bibinfo  {journal} {Physical review letters}\
  }\textbf {\bibinfo {volume} {110}},\ \bibinfo {pages} {153606} (\bibinfo
  {year} {2013})}\BibitemShut {NoStop}%
\bibitem [{\citenamefont {Roghani}\ and\ \citenamefont
  {Helm}(2008)}]{RoghaniHelm2008}%
  \BibitemOpen
  \bibfield  {author} {\bibinfo {author} {\bibfnamefont {M.}~\bibnamefont
  {Roghani}}\ and\ \bibinfo {author} {\bibfnamefont {H.}~\bibnamefont {Helm}},\
  }\href@noop {} {\bibfield  {journal} {\bibinfo  {journal} {Physical Review
  A}\ }\textbf {\bibinfo {volume} {77}},\ \bibinfo {pages} {043418} (\bibinfo
  {year} {2008})}\BibitemShut {NoStop}%
\bibitem [{\citenamefont {Joshi}\ \emph {et~al.}(2019)\citenamefont {Joshi},
  \citenamefont {Hrmo}, \citenamefont {Jarlaud}, \citenamefont {Oehl},\ and\
  \citenamefont {Thompson}}]{JoshiThompson2019}%
  \BibitemOpen
  \bibfield  {author} {\bibinfo {author} {\bibfnamefont {M.}~\bibnamefont
  {Joshi}}, \bibinfo {author} {\bibfnamefont {P.}~\bibnamefont {Hrmo}},
  \bibinfo {author} {\bibfnamefont {V.}~\bibnamefont {Jarlaud}}, \bibinfo
  {author} {\bibfnamefont {F.}~\bibnamefont {Oehl}}, \ and\ \bibinfo {author}
  {\bibfnamefont {R.}~\bibnamefont {Thompson}},\ }\href@noop {} {\bibfield
  {journal} {\bibinfo  {journal} {Physical Review A}\ }\textbf {\bibinfo
  {volume} {99}},\ \bibinfo {pages} {013423} (\bibinfo {year}
  {2019})}\BibitemShut {NoStop}%
\bibitem [{\citenamefont {Lindblad}(1976)}]{Lindblad1976}%
  \BibitemOpen
  \bibfield  {author} {\bibinfo {author} {\bibfnamefont {G.}~\bibnamefont
  {Lindblad}},\ }\href {\doibase 10.1007/BF01608499} {\bibfield  {journal}
  {\bibinfo  {journal} {Comm. Math. Phys.}\ }\textbf {\bibinfo {volume} {48}},\
  \bibinfo {pages} {119} (\bibinfo {year} {1976})}\BibitemShut {NoStop}%
\bibitem [{\citenamefont {Gorini}\ \emph {et~al.}(1976)\citenamefont {Gorini},
  \citenamefont {Kossakowski},\ and\ \citenamefont
  {Sudarshan}}]{GoriniSudarshan1976}%
  \BibitemOpen
  \bibfield  {author} {\bibinfo {author} {\bibfnamefont {V.}~\bibnamefont
  {Gorini}}, \bibinfo {author} {\bibfnamefont {A.}~\bibnamefont {Kossakowski}},
  \ and\ \bibinfo {author} {\bibfnamefont {E.~C.~G.}\ \bibnamefont
  {Sudarshan}},\ }\href {\doibase 10.1063/1.522979} {\bibfield  {journal}
  {\bibinfo  {journal} {J. Math. Phys.}\ }\textbf {\bibinfo {volume} {17}},\
  \bibinfo {pages} {821} (\bibinfo {year} {1976})}\BibitemShut {NoStop}%
\bibitem [{\citenamefont {Landi}\ \emph {et~al.}(2021)\citenamefont {Landi},
  \citenamefont {Poletti},\ and\ \citenamefont {Schaller}}]{LandiSchaller2021}%
  \BibitemOpen
  \bibfield  {author} {\bibinfo {author} {\bibfnamefont {G.~T.}\ \bibnamefont
  {Landi}}, \bibinfo {author} {\bibfnamefont {D.}~\bibnamefont {Poletti}}, \
  and\ \bibinfo {author} {\bibfnamefont {G.}~\bibnamefont {Schaller}},\
  }\href@noop {} {\bibfield  {journal} {\bibinfo  {journal} {arXiv preprint
  arXiv:2104.14350}\ } (\bibinfo {year} {2021})}\BibitemShut {NoStop}%
\bibitem [{\citenamefont {Guo}\ and\ \citenamefont
  {Poletti}(2021)}]{GuoPoletti2021}%
  \BibitemOpen
  \bibfield  {author} {\bibinfo {author} {\bibfnamefont {C.}~\bibnamefont
  {Guo}}\ and\ \bibinfo {author} {\bibfnamefont {D.}~\bibnamefont {Poletti}},\
  }\href@noop {} {\bibfield  {journal} {\bibinfo  {journal} {Physical Review
  E}\ }\textbf {\bibinfo {volume} {103}},\ \bibinfo {pages} {013309} (\bibinfo
  {year} {2021})}\BibitemShut {NoStop}%
\bibitem [{\citenamefont {Innes}\ \emph {et~al.}(2019)\citenamefont {Innes},
  \citenamefont {Edelman}, \citenamefont {Fischer}, \citenamefont {Rackauckus},
  \citenamefont {Saba}, \citenamefont {Shah},\ and\ \citenamefont
  {Tebbutt}}]{Zygote2019}%
  \BibitemOpen
  \bibfield  {author} {\bibinfo {author} {\bibfnamefont {M.}~\bibnamefont
  {Innes}}, \bibinfo {author} {\bibfnamefont {A.}~\bibnamefont {Edelman}},
  \bibinfo {author} {\bibfnamefont {K.}~\bibnamefont {Fischer}}, \bibinfo
  {author} {\bibfnamefont {C.}~\bibnamefont {Rackauckus}}, \bibinfo {author}
  {\bibfnamefont {E.}~\bibnamefont {Saba}}, \bibinfo {author} {\bibfnamefont
  {V.~B.}\ \bibnamefont {Shah}}, \ and\ \bibinfo {author} {\bibfnamefont
  {W.}~\bibnamefont {Tebbutt}},\ }\href@noop {} {\bibfield  {journal} {\bibinfo
   {journal} {arXiv:1907.07587}\ } (\bibinfo {year} {2019})}\BibitemShut
  {NoStop}%
\bibitem [{sou()}]{sourcecode}%
  \BibitemOpen
  \href@noop {} {\bibinfo  {journal}
  {https://github.com/workinghard-lyworking/Cooling-Optimal}\ }\BibitemShut
  {NoStop}%
\bibitem [{\citenamefont {Liu}\ and\ \citenamefont {Nocedal}(1989)}]{LBFGS}%
  \BibitemOpen
\bibfield  {journal} {  }\bibfield  {author} {\bibinfo {author} {\bibfnamefont
  {D.~C.}\ \bibnamefont {Liu}}\ and\ \bibinfo {author} {\bibfnamefont
  {J.}~\bibnamefont {Nocedal}},\ }\href@noop {} {\bibfield  {journal} {\bibinfo
   {journal} {Mathematical programming}\ }\textbf {\bibinfo {volume} {45}},\
  \bibinfo {pages} {503} (\bibinfo {year} {1989})}\BibitemShut {NoStop}%
\bibitem [{\citenamefont {Khaneja}\ \emph {et~al.}(2001)\citenamefont
  {Khaneja}, \citenamefont {Brockett},\ and\ \citenamefont
  {Glaser}}]{2001Time}%
  \BibitemOpen
  \bibfield  {author} {\bibinfo {author} {\bibfnamefont {N.}~\bibnamefont
  {Khaneja}}, \bibinfo {author} {\bibfnamefont {R.}~\bibnamefont {Brockett}}, \
  and\ \bibinfo {author} {\bibfnamefont {S.~J.}\ \bibnamefont {Glaser}},\
  }\href@noop {} {\bibfield  {journal} {\bibinfo  {journal} {Physical Review
  A}\ }\textbf {\bibinfo {volume} {63}},\ \bibinfo {pages} {032308} (\bibinfo
  {year} {2001})}\BibitemShut {NoStop}%
\bibitem [{\citenamefont {Schirmer}\ and\ \citenamefont
  {de~Fouquieres}(2011)}]{SchirmerFouquieres2011}%
  \BibitemOpen
  \bibfield  {author} {\bibinfo {author} {\bibfnamefont {S.~G.}\ \bibnamefont
  {Schirmer}}\ and\ \bibinfo {author} {\bibfnamefont {P.}~\bibnamefont
  {de~Fouquieres}},\ }\href@noop {} {\bibfield  {journal} {\bibinfo  {journal}
  {New Journal of Physics}\ }\textbf {\bibinfo {volume} {13}},\ \bibinfo
  {pages} {073029} (\bibinfo {year} {2011})}\BibitemShut {NoStop}%
\bibitem [{\citenamefont {{Machnes}}\ \emph {et~al.}(2011)\citenamefont
  {{Machnes}}, \citenamefont {{Sander}}, \citenamefont {{Glaser}},
  \citenamefont {{de Fouqui{\`e}res}}, \citenamefont {{Gruslys}}, \citenamefont
  {{Schirmer}},\ and\ \citenamefont
  {{Schulte-Herbr{\"u}ggen}}}]{2011Comparing}%
  \BibitemOpen
  \bibfield  {author} {\bibinfo {author} {\bibfnamefont {S.}~\bibnamefont
  {{Machnes}}}, \bibinfo {author} {\bibfnamefont {U.}~\bibnamefont {{Sander}}},
  \bibinfo {author} {\bibfnamefont {S.~J.}\ \bibnamefont {{Glaser}}}, \bibinfo
  {author} {\bibfnamefont {P.}~\bibnamefont {{de Fouqui{\`e}res}}}, \bibinfo
  {author} {\bibfnamefont {A.}~\bibnamefont {{Gruslys}}}, \bibinfo {author}
  {\bibfnamefont {S.}~\bibnamefont {{Schirmer}}}, \ and\ \bibinfo {author}
  {\bibfnamefont {T.}~\bibnamefont {{Schulte-Herbr{\"u}ggen}}},\ }\href
  {\doibase 10.1103/PhysRevA.84.022305} {\bibfield  {journal} {\bibinfo
  {journal} {\pra}\ }\textbf {\bibinfo {volume} {84}},\ \bibinfo {eid} {022305}
  (\bibinfo {year} {2011})},\ \Eprint {http://arxiv.org/abs/1011.4874}
  {arXiv:1011.4874 [quant-ph]} \BibitemShut {NoStop}%
\bibitem [{\citenamefont {{de Fouquieres}}\ \emph {et~al.}(2011)\citenamefont
  {{de Fouquieres}}, \citenamefont {Schirmer}, \citenamefont {Glaser},\ and\
  \citenamefont {Kuprov}}]{2011Second}%
  \BibitemOpen
  \bibfield  {author} {\bibinfo {author} {\bibfnamefont {P.}~\bibnamefont {{de
  Fouquieres}}}, \bibinfo {author} {\bibfnamefont {S.}~\bibnamefont
  {Schirmer}}, \bibinfo {author} {\bibfnamefont {S.}~\bibnamefont {Glaser}}, \
  and\ \bibinfo {author} {\bibfnamefont {I.}~\bibnamefont {Kuprov}},\ }\href
  {\doibase https://doi.org/10.1016/j.jmr.2011.07.023} {\bibfield  {journal}
  {\bibinfo  {journal} {Journal of Magnetic Resonance}\ }\textbf {\bibinfo
  {volume} {212}},\ \bibinfo {pages} {412} (\bibinfo {year}
  {2011})}\BibitemShut {NoStop}%
\bibitem [{\citenamefont {Floether}\ \emph {et~al.}(2012)\citenamefont
  {Floether}, \citenamefont {de~Fouquieres},\ and\ \citenamefont
  {Schirmer}}]{Floether_2012}%
  \BibitemOpen
  \bibfield  {author} {\bibinfo {author} {\bibfnamefont {F.~F.}\ \bibnamefont
  {Floether}}, \bibinfo {author} {\bibfnamefont {P.}~\bibnamefont
  {de~Fouquieres}}, \ and\ \bibinfo {author} {\bibfnamefont {S.~G.}\
  \bibnamefont {Schirmer}},\ }\href {\doibase 10.1088/1367-2630/14/7/073023}
  {\bibfield  {journal} {\bibinfo  {journal} {New Journal of Physics}\ }\textbf
  {\bibinfo {volume} {14}},\ \bibinfo {pages} {073023} (\bibinfo {year}
  {2012})}\BibitemShut {NoStop}%
\bibitem [{\citenamefont {Zahedinejad}\ \emph {et~al.}(2014)\citenamefont
  {Zahedinejad}, \citenamefont {Schirmer},\ and\ \citenamefont
  {Sanders}}]{2014Evolutionary}%
  \BibitemOpen
  \bibfield  {author} {\bibinfo {author} {\bibfnamefont {E.}~\bibnamefont
  {Zahedinejad}}, \bibinfo {author} {\bibfnamefont {S.}~\bibnamefont
  {Schirmer}}, \ and\ \bibinfo {author} {\bibfnamefont {B.~C.}\ \bibnamefont
  {Sanders}},\ }\href@noop {} {\bibfield  {journal} {\bibinfo  {journal}
  {Physical Review A}\ }\textbf {\bibinfo {volume} {90}},\ \bibinfo {pages}
  {032310} (\bibinfo {year} {2014})}\BibitemShut {NoStop}%
\bibitem [{\citenamefont {Riviello}\ \emph {et~al.}(2014)\citenamefont
  {Riviello}, \citenamefont {Brif}, \citenamefont {Long}, \citenamefont {Wu},
  \citenamefont {Tibbetts}, \citenamefont {Ho},\ and\ \citenamefont
  {Rabitz}}]{2014Searching}%
  \BibitemOpen
  \bibfield  {author} {\bibinfo {author} {\bibfnamefont {G.}~\bibnamefont
  {Riviello}}, \bibinfo {author} {\bibfnamefont {C.}~\bibnamefont {Brif}},
  \bibinfo {author} {\bibfnamefont {R.}~\bibnamefont {Long}}, \bibinfo {author}
  {\bibfnamefont {R.~B.}\ \bibnamefont {Wu}}, \bibinfo {author} {\bibfnamefont
  {K.~M.}\ \bibnamefont {Tibbetts}}, \bibinfo {author} {\bibfnamefont {T.~S.}\
  \bibnamefont {Ho}}, \ and\ \bibinfo {author} {\bibfnamefont {H.}~\bibnamefont
  {Rabitz}},\ }\href@noop {} {\bibfield  {journal} {\bibinfo  {journal}
  {Physical Review A}\ }\textbf {\bibinfo {volume} {90}},\ \bibinfo {pages}
  {013404} (\bibinfo {year} {2014})}\BibitemShut {NoStop}%
\bibitem [{\citenamefont {Jensen}\ \emph {et~al.}(2020)\citenamefont {Jensen},
  \citenamefont {M{\o}ller}, \citenamefont {S{\o}rensen},\ and\ \citenamefont
  {Sherson}}]{JensenSherson2020}%
  \BibitemOpen
  \bibfield  {author} {\bibinfo {author} {\bibfnamefont {J.~H.~M.}\
  \bibnamefont {Jensen}}, \bibinfo {author} {\bibfnamefont {F.~S.}\
  \bibnamefont {M{\o}ller}}, \bibinfo {author} {\bibfnamefont {J.~J.}\
  \bibnamefont {S{\o}rensen}}, \ and\ \bibinfo {author} {\bibfnamefont {J.~F.}\
  \bibnamefont {Sherson}},\ }\href@noop {} {\bibfield  {journal} {\bibinfo
  {journal} {arXiv preprint arXiv:2005.09943}\ } (\bibinfo {year}
  {2020})}\BibitemShut {NoStop}%
\bibitem [{\citenamefont {Sch{\"a}fer}\ \emph {et~al.}(2020)\citenamefont
  {Sch{\"a}fer}, \citenamefont {Kloc}, \citenamefont {Bruder},\ and\
  \citenamefont {L{\"o}rch}}]{SchaferLorch2020}%
  \BibitemOpen
  \bibfield  {author} {\bibinfo {author} {\bibfnamefont {F.}~\bibnamefont
  {Sch{\"a}fer}}, \bibinfo {author} {\bibfnamefont {M.}~\bibnamefont {Kloc}},
  \bibinfo {author} {\bibfnamefont {C.}~\bibnamefont {Bruder}}, \ and\ \bibinfo
  {author} {\bibfnamefont {N.}~\bibnamefont {L{\"o}rch}},\ }\href@noop {}
  {\bibfield  {journal} {\bibinfo  {journal} {Machine Learning: Science and
  Technology}\ }\textbf {\bibinfo {volume} {1}},\ \bibinfo {pages} {035009}
  (\bibinfo {year} {2020})}\BibitemShut {NoStop}%
\bibitem [{\citenamefont {Vargas-Hern{\'a}ndez}\ \emph
  {et~al.}(2021)\citenamefont {Vargas-Hern{\'a}ndez}, \citenamefont {Chen},
  \citenamefont {Jung},\ and\ \citenamefont {Brumer}}]{HernandezBrumer2021}%
  \BibitemOpen
  \bibfield  {author} {\bibinfo {author} {\bibfnamefont {R.~A.}\ \bibnamefont
  {Vargas-Hern{\'a}ndez}}, \bibinfo {author} {\bibfnamefont {R.~T.}\
  \bibnamefont {Chen}}, \bibinfo {author} {\bibfnamefont {K.~A.}\ \bibnamefont
  {Jung}}, \ and\ \bibinfo {author} {\bibfnamefont {P.}~\bibnamefont
  {Brumer}},\ }\href@noop {} {\bibfield  {journal} {\bibinfo  {journal} {arXiv
  preprint arXiv:2103.12604}\ } (\bibinfo {year} {2021})}\BibitemShut {NoStop}%
\bibitem [{\citenamefont {Marzoli}\ \emph {et~al.}(1994)\citenamefont
  {Marzoli}, \citenamefont {Cirac}, \citenamefont {Blatt},\ and\ \citenamefont
  {Zoller}}]{MarzoliZoller1994}%
  \BibitemOpen
  \bibfield  {author} {\bibinfo {author} {\bibfnamefont {I.}~\bibnamefont
  {Marzoli}}, \bibinfo {author} {\bibfnamefont {J.}~\bibnamefont {Cirac}},
  \bibinfo {author} {\bibfnamefont {R.}~\bibnamefont {Blatt}}, \ and\ \bibinfo
  {author} {\bibfnamefont {P.}~\bibnamefont {Zoller}},\ }\href@noop {}
  {\bibfield  {journal} {\bibinfo  {journal} {Physical Review A}\ }\textbf
  {\bibinfo {volume} {49}},\ \bibinfo {pages} {2771} (\bibinfo {year}
  {1994})}\BibitemShut {NoStop}%
\bibitem [{\citenamefont {Roos}(2000)}]{roosthesis}%
  \BibitemOpen
  \bibfield  {author} {\bibinfo {author} {\bibfnamefont {C.}~\bibnamefont
  {Roos}},\ }\emph {\bibinfo {title} {Controlling the quantum state of trapped
  ions}},\ \href@noop {} {Ph.D. thesis} (\bibinfo {year} {2000})\BibitemShut
  {NoStop}%
\bibitem [{\citenamefont {Morigi}(2003)}]{Morigi2003}%
  \BibitemOpen
  \bibfield  {author} {\bibinfo {author} {\bibfnamefont {G.}~\bibnamefont
  {Morigi}},\ }\href@noop {} {\bibfield  {journal} {\bibinfo  {journal}
  {Physical Review A}\ }\textbf {\bibinfo {volume} {67}},\ \bibinfo {pages}
  {033402} (\bibinfo {year} {2003})}\BibitemShut {NoStop}%
\bibitem [{\citenamefont {Rei}\ \emph {et~al.}(1996)\citenamefont {Rei},
  \citenamefont {Lindner}, \citenamefont {Blatt} \emph
  {et~al.}}]{ReibBlatt1996}%
  \BibitemOpen
  \bibfield  {author} {\bibinfo {author} {\bibfnamefont {D.}~\bibnamefont
  {Rei}}, \bibinfo {author} {\bibfnamefont {A.}~\bibnamefont {Lindner}},
  \bibinfo {author} {\bibfnamefont {R.}~\bibnamefont {Blatt}},  \emph
  {et~al.},\ }\href@noop {} {\bibfield  {journal} {\bibinfo  {journal}
  {Physical Review A}\ }\textbf {\bibinfo {volume} {54}},\ \bibinfo {pages}
  {5133} (\bibinfo {year} {1996})}\BibitemShut {NoStop}%
\bibitem [{\citenamefont {Thompson}\ \emph {et~al.}(2013)\citenamefont
  {Thompson}, \citenamefont {Tiecke}, \citenamefont {Zibrov}, \citenamefont
  {Vuleti{\'c}},\ and\ \citenamefont {Lukin}}]{Thompson2013coherence}%
  \BibitemOpen
  \bibfield  {author} {\bibinfo {author} {\bibfnamefont {J.~D.}\ \bibnamefont
  {Thompson}}, \bibinfo {author} {\bibfnamefont {T.}~\bibnamefont {Tiecke}},
  \bibinfo {author} {\bibfnamefont {A.~S.}\ \bibnamefont {Zibrov}}, \bibinfo
  {author} {\bibfnamefont {V.}~\bibnamefont {Vuleti{\'c}}}, \ and\ \bibinfo
  {author} {\bibfnamefont {M.~D.}\ \bibnamefont {Lukin}},\ }\href@noop {}
  {\bibfield  {journal} {\bibinfo  {journal} {Physical review letters}\
  }\textbf {\bibinfo {volume} {110}},\ \bibinfo {pages} {133001} (\bibinfo
  {year} {2013})}\BibitemShut {NoStop}%
\bibitem [{\citenamefont {Kaufman}\ \emph {et~al.}(2012)\citenamefont
  {Kaufman}, \citenamefont {Lester},\ and\ \citenamefont
  {Regal}}]{Kaufman2012cooling}%
  \BibitemOpen
  \bibfield  {author} {\bibinfo {author} {\bibfnamefont {A.~M.}\ \bibnamefont
  {Kaufman}}, \bibinfo {author} {\bibfnamefont {B.~J.}\ \bibnamefont {Lester}},
  \ and\ \bibinfo {author} {\bibfnamefont {C.~A.}\ \bibnamefont {Regal}},\
  }\href@noop {} {\bibfield  {journal} {\bibinfo  {journal} {Physical Review
  X}\ }\textbf {\bibinfo {volume} {2}},\ \bibinfo {pages} {041014} (\bibinfo
  {year} {2012})}\BibitemShut {NoStop}%
\bibitem [{\citenamefont {Reiserer}\ \emph {et~al.}(2013)\citenamefont
  {Reiserer}, \citenamefont {N{\"o}lleke}, \citenamefont {Ritter},\ and\
  \citenamefont {Rempe}}]{reiserer2013ground}%
  \BibitemOpen
  \bibfield  {author} {\bibinfo {author} {\bibfnamefont {A.}~\bibnamefont
  {Reiserer}}, \bibinfo {author} {\bibfnamefont {C.}~\bibnamefont
  {N{\"o}lleke}}, \bibinfo {author} {\bibfnamefont {S.}~\bibnamefont {Ritter}},
  \ and\ \bibinfo {author} {\bibfnamefont {G.}~\bibnamefont {Rempe}},\
  }\href@noop {} {\bibfield  {journal} {\bibinfo  {journal} {Physical review
  letters}\ }\textbf {\bibinfo {volume} {110}},\ \bibinfo {pages} {223003}
  (\bibinfo {year} {2013})}\BibitemShut {NoStop}%
\bibitem [{\citenamefont {Chan}\ \emph {et~al.}(2011)\citenamefont {Chan},
  \citenamefont {Alegre}, \citenamefont {Safavi-Naeini}, \citenamefont {Hill},
  \citenamefont {Krause}, \citenamefont {Gr{\"o}blacher}, \citenamefont
  {Aspelmeyer},\ and\ \citenamefont {Painter}}]{chan2011laser}%
  \BibitemOpen
  \bibfield  {author} {\bibinfo {author} {\bibfnamefont {J.}~\bibnamefont
  {Chan}}, \bibinfo {author} {\bibfnamefont {T.~M.}\ \bibnamefont {Alegre}},
  \bibinfo {author} {\bibfnamefont {A.~H.}\ \bibnamefont {Safavi-Naeini}},
  \bibinfo {author} {\bibfnamefont {J.~T.}\ \bibnamefont {Hill}}, \bibinfo
  {author} {\bibfnamefont {A.}~\bibnamefont {Krause}}, \bibinfo {author}
  {\bibfnamefont {S.}~\bibnamefont {Gr{\"o}blacher}}, \bibinfo {author}
  {\bibfnamefont {M.}~\bibnamefont {Aspelmeyer}}, \ and\ \bibinfo {author}
  {\bibfnamefont {O.}~\bibnamefont {Painter}},\ }\href@noop {} {\bibfield
  {journal} {\bibinfo  {journal} {Nature}\ }\textbf {\bibinfo {volume} {478}},\
  \bibinfo {pages} {89} (\bibinfo {year} {2011})}\BibitemShut {NoStop}%
\end{thebibliography}

\end{document}